\documentclass{pnastwo}

\usepackage{graphicx}
\usepackage{overpic}
\usepackage{amssymb,amsfonts,amsmath}

\copyrightyear{2008}
\issuedate{Issue Date}
\volume{Volume}
\issuenumber{Issue Number}

\begin{document}

\title{The expert game --- Cooperation in social communication.}

\author{Kristian Moss Bendtsen\affil{1}{Center for Models of Life, Niels Bohr Institute, University of Copenhagen, Blegdamsvej 17, DK-2100, Copenhagen, Denmark},
Florian Uekermann\affil{1}{},
\and
Jan O. Haerter\affil{1}{}}

\contributor{Manuscript in preparation for submission.}

\maketitle

\begin{article}
\begin{abstract}
Large parts of professional human communication proceed in a request-reply fashion, whereby requests contain specifics of the information desired while replies can deliver the required information. 
However, time limitations often force individuals to prioritize some while neglecting others.
This dilemma will inevitably force individuals into defecting against some communication partners to give attention to others. 
Furthermore, communication entirely breaks down when individuals act purely egoistically as replies would never be issued and quest for desired information would always be prioritized.
Here we present an experiment, termed ``The expert game'', where a number of individuals communicate with one-another through an electronic messaging system. 
By imposing a strict limit on the number of sent messages, individuals were required to decide between requesting information that is beneficial for themselves or helping others by replying to their requests.
In the experiment, individuals were assigned the task to find the expert on a specific topic and receive a reply from that expert.
Tasks and expertise of each player were periodically re-assigned to randomize the required interactions.
Resisting this randomization, a non-random network of cooperative communication between individuals formed. 
We use a simple Bayesian inference algorithm to model each player's trust in the cooperativity of others with good experimental agreement.
Our results suggest that human communication in groups of individuals is strategic and favors cooperation with trusted parties at the cost of defection against others. 
To establish and maintain trusted links a significant fraction of time-resources is allocated, even in situations where the information transmitted is negligible. 
\end{abstract}


\dropcap{B}eyond the notion of kin selection, achieving cooperation between individuals has long appeared as a riddle in evolution \cite{rand2013human}. 
Yet, theories in favor of cooperation in repeated interactions involving memory are now well-established \cite{axelrod1981evolution, nowak2004emergence}.
When described as iterated Prisoner's Dilemmas, strategies such as {\it tit-for-tat} (TFT), a form of {\it direct reciprocity}, can out-compete repeated defection and a group of defectors can be invaded by a new individual playing TFT.
In support of these theories, empirical evidence for different types of cooperative strategies have been presented \cite{rand2013human}.
Some of this evidence is based on experiments with human subjects, in an effort to study cooperation in human society.  
It was discovered that the details of the connectedness of subjects are crucial for the establishment of cooperation. 
When subjects are organized on a regular lattice with predefined links to neighboring subjects, initial cooperation eventually drops to the same low levels that are obtained when links are randomly reshuffled after every repetition of the game \cite{grujic2010social}. These findings were supported by theoretical models \cite{gracia2012heterogeneous}. 
Conversely, cooperation has been shown to emerge if individuals can choose their collaborators. In this case cooperators cut links with notorious defectors \cite{rand2011dynamic}, leading to their isolation in the network.

In the complex social network community, the importance of dynamically evolving network structure has long been recognized \cite{gross2008adaptive}. 
Today's society --- especially with regards to business, academic or government institutions --- is highly dependent on efficient communication, much of which takes place through online messaging systems, such as electronic mail and other messaging services.
With the ever-growing amount of highly resolved data from digital communication, the analysis of such data has correspondingly become a field of heavy research.
But how does cooperation in communication networks emerge and how is it maintained?
Theoretical efforts describing communication on networks exist \cite{rosvall2003modeling,marsili2004rise,sneppen2005hide,rosvall2005searchability,rosvall2006modeling,rosvall2009reinforced}, but little attention has been devoted to the establishment and reinforcement of cooperation in these networks.
One difficulty with electronic communication without face-to-face contact can stem from limited trust in the cooperativity of communication partners \cite{rocco1998trust}.
In experiments which mimicked collaborative tasks through electronic communication subjects were found to break agreements more frequently than if face-to-face contact was allowed.
The importance of informal communication in {\it virtual teams} was pointed out when cooperation takes place through electronic mail, phone, or video-conferences only \cite{hertel2005managing}.
In addition, communication is prone to defection, due to the inherent capacity limit imposed by time limitations: Input received in terms of electronic mail or other types of asynchronous messages simply cannot always be answered within the time available to subjects \cite{miritello2013limited,haerter2012communication}. 
Unresponsiveness may therefore be an intrinsic and often unavoidable form of defection in real-world communication.

Our objective is to determine, if even under conditions that principally favor selfish behavior, cooperation can evolve in communication networks.
We therefore remove all the influence of external factors that could promote cooperation, such as reputation through common knowledge of subjects' behavior --- a mechanism which is thought to lead to indirect reciprocity \cite{nowak2005evolution,engelmann2009indirect}. 
Further, we want to avoid the effect of emotional influence between subjects.
To mimic a particularly impersonal communication network, correspondence between individuals is constrained in such a way as to disallow all emotional content of messages. Messages are standardised and purvey only factual information and no expression of emotions, such as regret, thankfulness or anger.

Communication through electronic messages is typically asynchronous, i.e. there is a time-delay between sending a message until it is read and potentially replied to. 
Asynchronous communication places additional burden on efficiency and establishment of trust, e.g. when compared to telephone communication, which is inherently more personal and synchronous, since it demands immediate replies. 
In fact, in virtual teams, the role of responses in building trust has been recognized \cite{jarvenpaa1998communication}. 
This study find that a response is a means of signaling one's involvement, the willingness to understand another person's request and invest time and energy in finding a solution.
In short, a response creates trust and this may be independent of the type of response.
Trust is a concept inseparably tied to risk \cite{deutsch1958trust}. 
It has even been argued that trust may not be possible in global virtual teams \cite{handy1995trust}.
In this experiment we study how channels of cooperative communication can self-organize in an impersonal communication network, even if replying conflicts fundamentally with the subjects' immediate interests.

\subsection{Information flow in asynchronous communication}
In asynchronous communication, obtaining specific, solicited information from others requires them to respond to a request for that information. 
A priori, an answer may appear as an act of altruism if it has a cost: There is no immediate payoff for its sender.
Thus it may be viewed as paradoxical, that this kind of communication can result in cooperation, especially under strict time constraints.

Imagine a group of professional individuals, where individuals in the group are operating under the pressure to complete a task under time constraints. 
However, each individual depends on the help of an expert to complete this task. 
A request for help with a task could be exemplified as follows:

Example of a request, A: {\it ``I am designing an interface for customers to order products on the internet. I have finished the artwork but am unable to carry out the programming.
Can you help me with that or do you know someone who can?''}

Note that the request contains both information on A's expertise and on the desired, lacking, expertise. 
In case the receiver of the request ``B'' does not have the expertise to help ``A'' complete the task, but knows someone who can, a possible response could be:

Example of a reply, B: {\it ``Yes, I do know someone and this person is C. Here is C's email address.''}

B's reply in the example contains information useful to A.
However ``B'' has to sacrifice valuable time to provide that information. 
Instead, ``B'' could spend this time looking for information valuable to ``B'' by sending requests to others.
To mimic this type of interaction, we propose an game termed {\it The expert game}.

\section{Experiment and model}
\subsection{Definition of the expert game}
\begin{center}
\begin{figure*}[t]
\begin{center}
\begin{overpic}[width=5cm,angle=-90,trim= 0cm 0pt 0pt 0pt,clip]{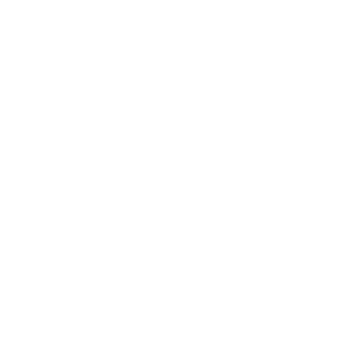}
\put(-150,-120){\includegraphics[width=17cm,angle=0]{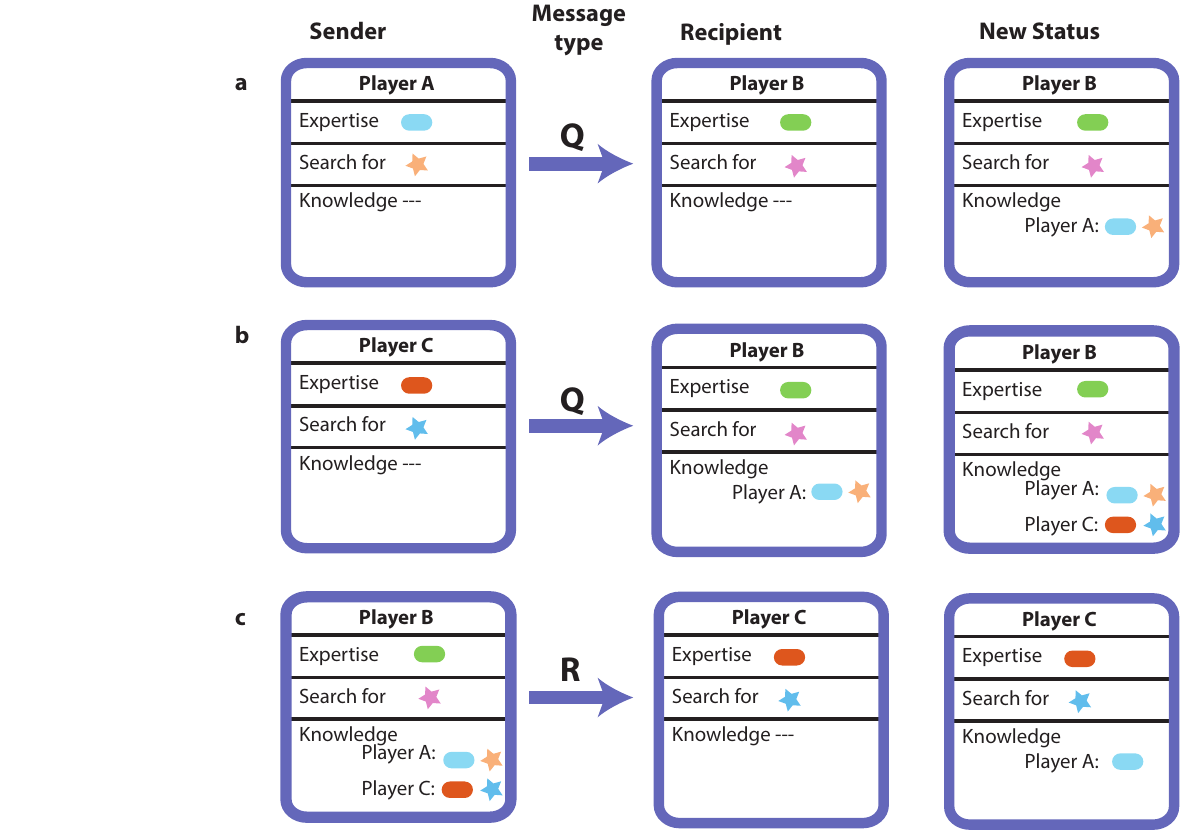}}
\end{overpic}
\vspace{180pt}
\caption{{\bf Expert game example.}  
Schematic showing a possible interaction between individuals. Matching colors denote matching expertise.
{\bf a}, Player A contacts player B in search of the expert. 
B thereby gains information both on the search target and the expertise of A.
{\bf b}, B receives another request by player C. 
B is able to match C's search request with previously gained information on A.
{\bf c}, B issues a positive reply (R) to C and directs C to its desired expert A.
C now has all relevant information and can contact A for help.
}
\label{fig:expert_game_schematic}
\end{center}
\end{figure*}
\end{center}

The expert game is a discrete game for $N>2$ participants.
In a game with $N$ participants, there are exactly $N$ different types of expertise and $N$ different tasks.
Each task matches one type of expertise, which is needed to complete the task.
Prior to the game, each participant is assigned a unique expertise and a unique task at random.
The only condition is that each expertise and the matching task are assigned to different participants.
Assigned expertise and task are revealed only to the participant they are assigned to.
Participants gain a point if they find the expert matching their task, send the expert a request for help and receive a reply to their request.
In each round each player can send at most one message to another player. 
There are four types of messages:
\begin{description}
\item[Request ({\it Q})] Can be sent to any other participant. The sender's expertise and task are revealed to the receiver.
\item[Reply] Participants can only send a reply to recipients who have previously sent them a request. 
There are three types of replies:
\begin{description}
\item[Confirmation ({\it C})] Can only be sent if the sender's expertise matches the receiver's task. The receiver's task is thereby successfully completed and he earns a point.
\item[Referral ({\it R})] Can only be sent if the sender's expertise does not match the receiver's task and if the sender knows whose expertise matches this task. The identity of the expert is revealed to the receiver.
\item[Negation ({\it N})] Can only be sent if the sender does not know whose expertise matches the receiver's task.
\end{description}
\end{description}
Subjects can also refuse to send a message altogether.
At the end of each round, when all participants have decided upon one of the above actions, all messages are simultaneously delivered.
An example of a possible interaction of players is shown in Fig.~\ref{fig:expert_game_schematic}.
Even when participants have earned a point (i.e. received a confirmation), they continue playing until the game is completed but they can not earn additional points during this game. 
The game is terminated after a certain number of rounds. The approximate number of rounds for each game is known to the participants but is varied randomly by small deviations to reduce edge effects.

The rules stated above allow participants to refuse sending messages, but any kind of deception is prohibited. 
The participants can therefore never choose between different replies, but are always restricted to reply with either a Confirmation (as the expert matching a corresponding request for expertise), a Referral (when not the expert but with knowledge on the expert's identity) or a Negation (no knowledge of the expert's identity).

\subsection{Experiment}
We performed a preliminary experiment with 4 consecutive games and 8 participants from the Biocomplexity group of the Niels Bohr Institute at the University of Copenhagen. 
Experiments with more participants in a more controlled setting are under preparation.
All participants were gathered in the same room but were instructed only to communicate with each other by means of a instant messaging interface which identified each participant by a virtual identity of a different name.
Expertise and tasks were represented by numbers, which were handed to the participants before each game.

\subsection{Model: Bayesian agents}\label{sec:model}
We model human behaviour in an iterated expert game using an agent-based model. 
In this model, the agents (i.e. participants) make individual estimates on each of the other agents' response probability.
Bayesian inference is applied to adjust the estimates of response probabilities. 
Re-assessment of probabilities is done between any two games, i.e. the data collected during a given game is employed to derive a posterior probability distribution from the previous prior. 
We further assign three constant personality traits, common to all agents.
Using these traits along with the estimated responsiveness, agents choose message type and communication partner in each round.

{\bf Responsiveness estimate ---} The responsiveness $\theta$ represents the probability that a request is responded to in the next round. 
Assuming that this probability does not change, this constitutes a Bernoulli process. 
The likelihood that a hypothesized $\theta$ leads to a reply in round $k$ after the request therefore matches the waiting time distribution for the first success in a Bernoulli process. The likelihood that a game ends before a reply is received is the probability of not succeeding once in $k$ Bernoulli trials, where $k$ represents the number of rounds that were played after the request was sent:
\begin{figure*}[h]
\begin{align}
\label{eq: waitingTimeDist} P(k|\theta)=
\begin{cases}
(1-\theta)^{k-1}\cdot\theta &\text{Reply received in round k since the request}\\
(1-\theta)^k &\text{No answer, game ended k rounds after request}
\end{cases}
\end{align}
\end{figure*}
In our model, each agent $x$ maintains a set of hypotheses of every other agent $y$'s responsiveness $\theta_{xy}$. 
These hypotheses are then used as a basis for choosing communication partners. 
The set of hypotheses used is $\theta_{xy}=0.05\cdot n$, with $0\leq\theta_{xy}\leq1$ and $n$ a non-negative integer. 
Each hypothesis is assigned an estimated probability  $P(\theta_{xy})$ of being correct. 
After each game the estimated probability for each hypothesis $\theta_{xy}$ is updated if $x$ sent a request to $y$. The updated probability estimates considering the latest reply time are determined using Bayesian inference:
\begin{align}
P(\theta_{xy}|k)\propto P(k|\theta_{xy})\cdot P(\theta_{xy}) \qquad\text{with}\qquad \sum\limits_{\theta_{xy}}P(\theta_{xy})=1\;.
\end{align}
Here, $P(\theta_{xy}|k)$ represents the updated probability estimate of the hypothesis that the responsiveness is $\theta_{xy}$, given the latest reply time $k$ as evidence.
$P(k|\theta_{xy})$ represents the likelihood of the observed reply time given the hypothesis $\theta_{xy}$ as defined in equation \ref{eq: waitingTimeDist}, and $P(\theta_{xy})$ represents the prior probability estimate of hypothesis $\theta_{xy}$. The updated distribution becomes the prior distribution in the next game.

Only the first request to a particular agent and the first reply are taken into account. 
In the special, and generally rare, case that a referral (R) is sent after a negative reply (N) has been sent, this is considered to be an instant reply ($k=1$), regardless of the actual reply time.
This is to say that the second reply (R) is considered as a generous act valued similarly as an immediate reply.

Agent $x$ estimates agent $y$'s responsiveness by the mean reply probability $\langle\theta_{xy}\rangle$. This is given by the weighted average of the probability estimates of all hypotheses:
\begin{align}
\langle\theta_{xy}\rangle=\sum\limits_{\theta_{xy}}\theta_{xy} \cdot P(\theta_{xy}) \label{meanReplyProb}
\end{align}
This responsiveness estimate is used by $x$ as an estimated value of $y$ as a communication partner.

{\bf Choice of message type and communication partner ---}

In addition to the constraints given by the rules of the game, the modeled agents never send more than one message of each type to a particular agent during the same game. 
This is a point where the model may remove some of the actual human behavior: In reality, the experimental subjects do occasionally repeat similar messages.
Finally, agents will always instantly send a request to the expert for their task if they learn who this expert is and have not sent this expert a request previously. 

We characterize the agents' personalities in terms of their preference for replying to requests versus that of sending requests themselves. 
This is modeled using three constant personality traits: The preference to send a request ($\alpha$); the preference to send referrals and confirmations ($\beta$); the preference to send negations ($\gamma$), where we have for simplicity assumed the probabilities of referrals and confirmations to be similar.
The choice of message to be sent in a round is made at random from all possible messages meeting the criteria defined above. 
The probability of choosing a particular message is proportional to the senders' preference to send this type of message multiplied with their estimate of the receiver's responsiveness $\langle\theta_{xy}\rangle$  (see Appendix equation \ref{eq: choiceProb} for a more formalized expression). 
This is to say that a sender will make an estimate on the expected future payoff of sending a message to a particular recipient.
After learning who the expert for their task is, an agent's preference to send requests ($\alpha$) to any other participant than their expert is lowered to $0.1\%$ of its original value for the rest of the game.

\section{Results}
{\bf Experimental results ---} 
Generally, all players issued all types of messages. Approximately 66 percent of messages were requests (Q), 18 percent were referrals and confirmations (R\& C), and 15 percent were negations (N). This means, that most time-resources were in-fact allocated to the acquisition of information (Q), but a significant fraction was devoted to replies. 
Considering that referrals and confirmations purvey potentially $N$ times as much information as negations, the comparably large rate of negations is noteworthy. 

Table \ref{tab:summary_table} shows the time lag between requests and replies for requests that did not go unanswered.
The data show that there is an expected delay of replies of more than one round. This delay is somewhat longer for negative replies.
It also shows that the rate of sending positive replies is somewhat higher than that of negative replies.
More than half ($55\;\%$) of requests remained unanswered by subjects at the end of the game, even if the subjects did possess the knowledge requested.

How does information pass from one subject to another?
To explore this, we map the network formed by messages between subjects.
The adjacency matrices shown in the figure depict a fairly noisy pattern for games 1 to 4  (Fig.~\ref{fig:interaction_networks}b---e). 
However, one notable aspect of these matrices is that the first game has produced much more uniform sampling of all interactions.
In games 2 to 4 a more heterogeneous pattern is observed, with many messages between certain subjects, while other pairs of nodes do not interact at all. 
This can be interpreted as communication pathways forming at an early stage of the experiment and stabilizing, or evolving at a lower rate, later on. 

\begin{center}
\begin{figure*}[t]
\begin{center}
\begin{overpic}[width=5cm,angle=-90,trim= 0cm 0pt 0pt 0pt,clip]{dummy.pdf}
\put(-80,-20){\includegraphics[width=12.4cm,angle=0]{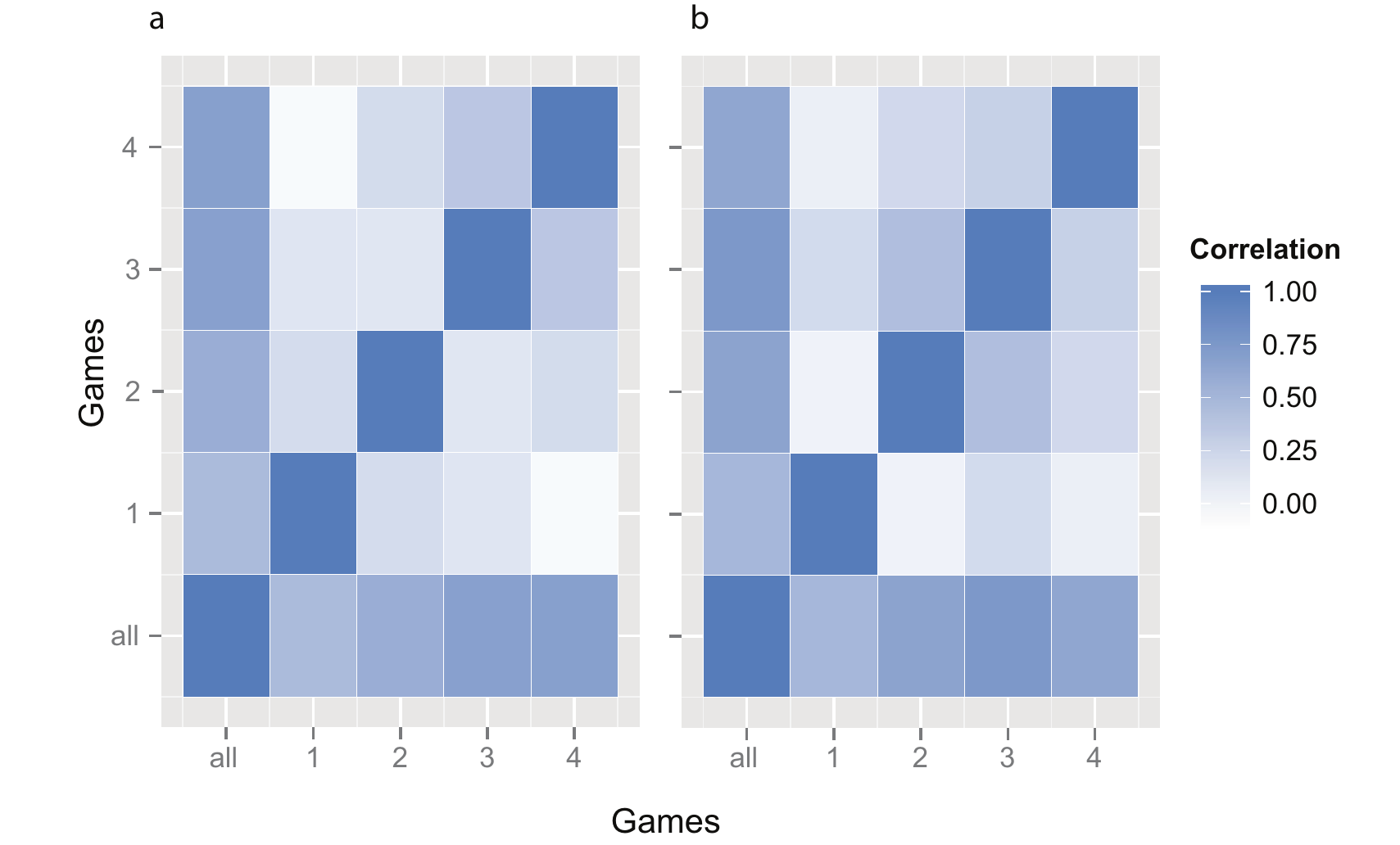}}
\end{overpic}
\vspace{50pt}
\caption{{\bf Correlation matrix.}  
{\bf a}, Experimental correlation between the adjacency matrices of different games. 
{\bf b}, Corresponding modeled correlation matrix.
}
\label{fig:correlation_matrix}
\end{center}
\end{figure*}
\end{center}

When condensing all communication into a single adjacency matrix (Fig.~\ref{fig:interaction_networks}a), a clear pattern of reciprocity emerges.
This is seen by the fact that the matrix is much more symmetric than would be expected from random interactions between nodes.
Reciprocity has in fact been recognized as a consistent feature of human communication and cooperation more generally \cite{fehr2003nature}.
To further explore the subjects' perception of their interactions, after the experiment, all subjects were asked to provide an ordered list of collaborators, ranging from best to worst.
From this list we composed the perceived adjacency matrix shown in Fig.~\ref{fig:interaction_networks}f. 
Note the qualitative similarity to Fig.~\ref{fig:interaction_networks}a.
This shows that the participants were aware of their preferred communication channels, a further confirmation that the reciprocal structure is far from random.
To assess if the networks of communication formed within the separate games are related, we compute the correlation coefficient between all combinations of any two networks (Fig.~\ref{fig:correlation_matrix}a).
The data shows that indeed positive correlations exist and are stronger the shorter the time between two games. 
The correlations between distinct games are strongest for the last and second to last game, indicating that communication channels may become more stable over time, i.e. mutual agreements to communicate and cooperate are re-enforced.

\begin{center}
\begin{figure*}[t]
\begin{center}
\begin{overpic}[width=5cm,angle=-90,trim= 0cm 0pt 0pt 0pt,clip]{dummy.pdf}
\put(-80,170){\includegraphics[height=6.4cm,angle=-90,trim= 0cm 0cm 1.4cm 0pt,clip]{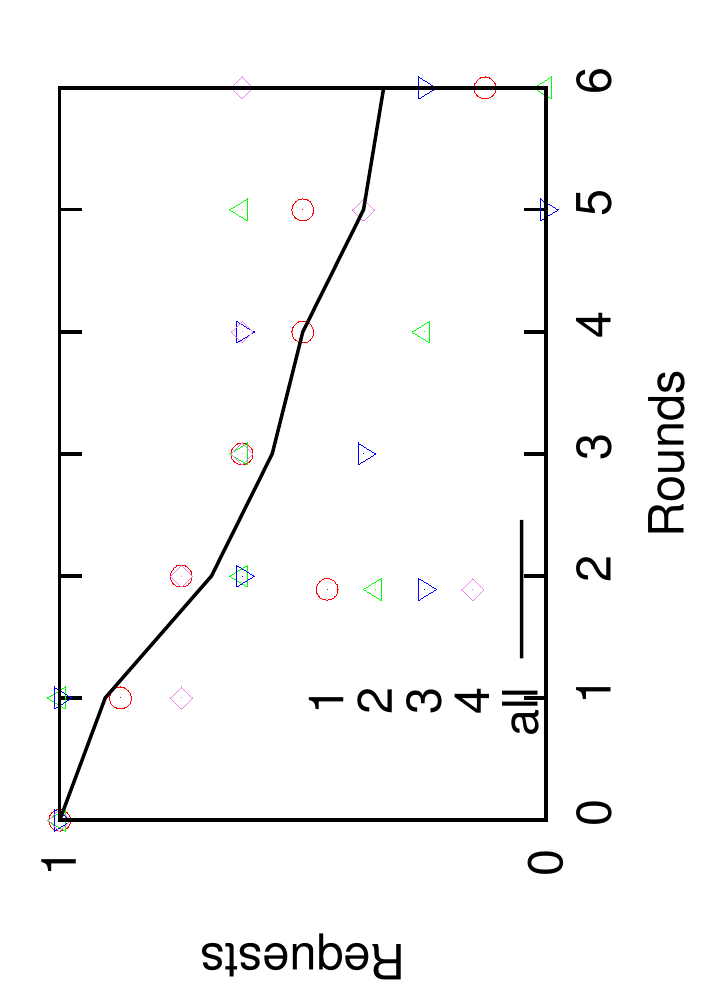}}
\put(-80,100){\includegraphics[height=6.4cm,angle=-90,trim= 0cm 0cm 1.4cm 0pt,clip]{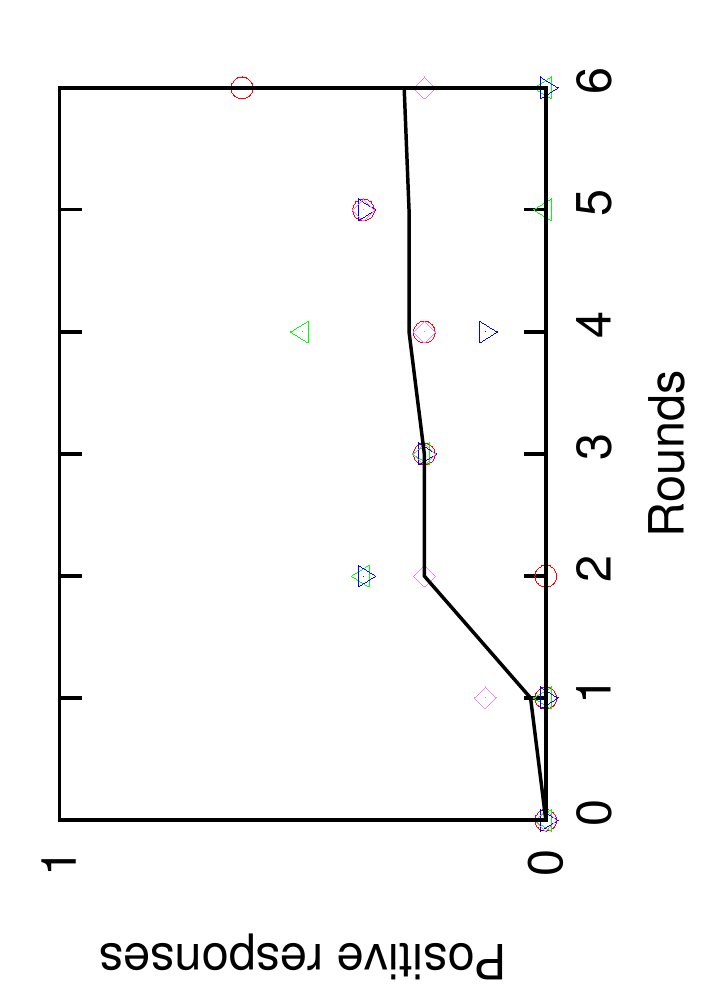}}
\put(-80,30){\includegraphics[height=6.4cm,angle=-90,trim= 0cm 0cm 1.4cm 0pt,clip]{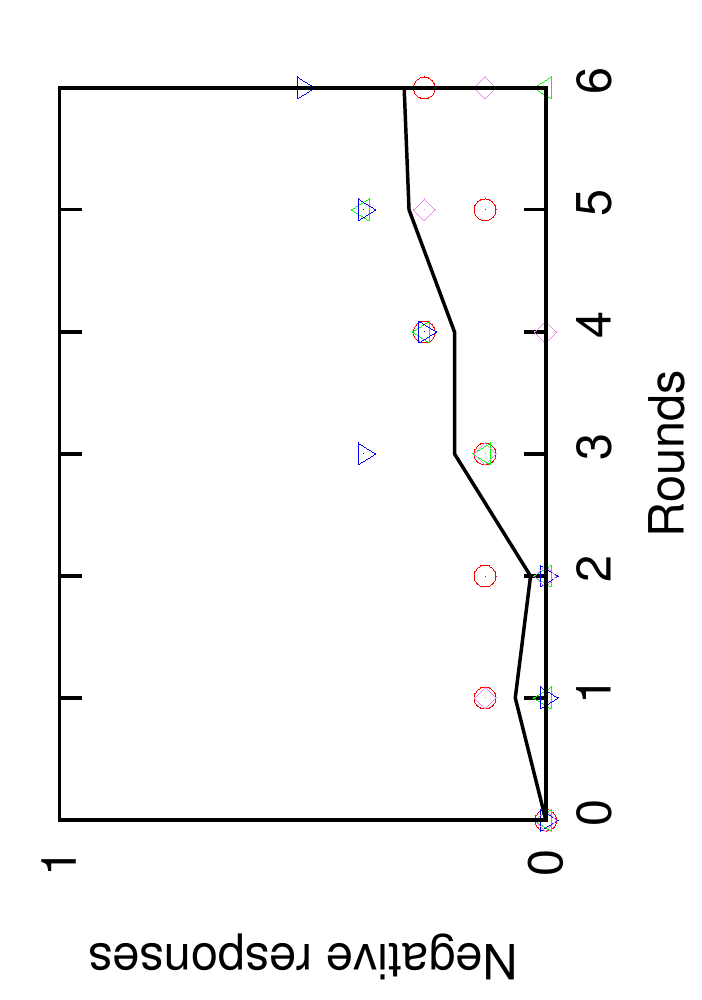}}
\put(-80,-40){\includegraphics[height=6.4cm,angle=-90,trim= 0cm 0cm 1.4cm 0pt,clip]{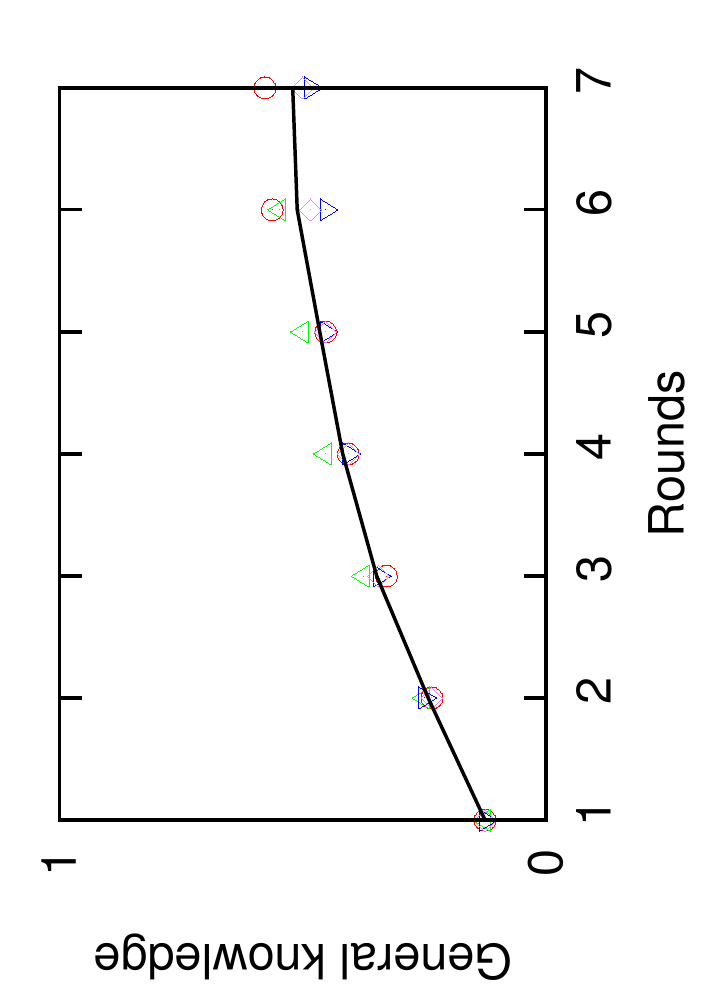}}
\put(-80,-110){\includegraphics[height=6.4cm,angle=-90,trim= 0cm 0cm 0cm 0pt,clip]{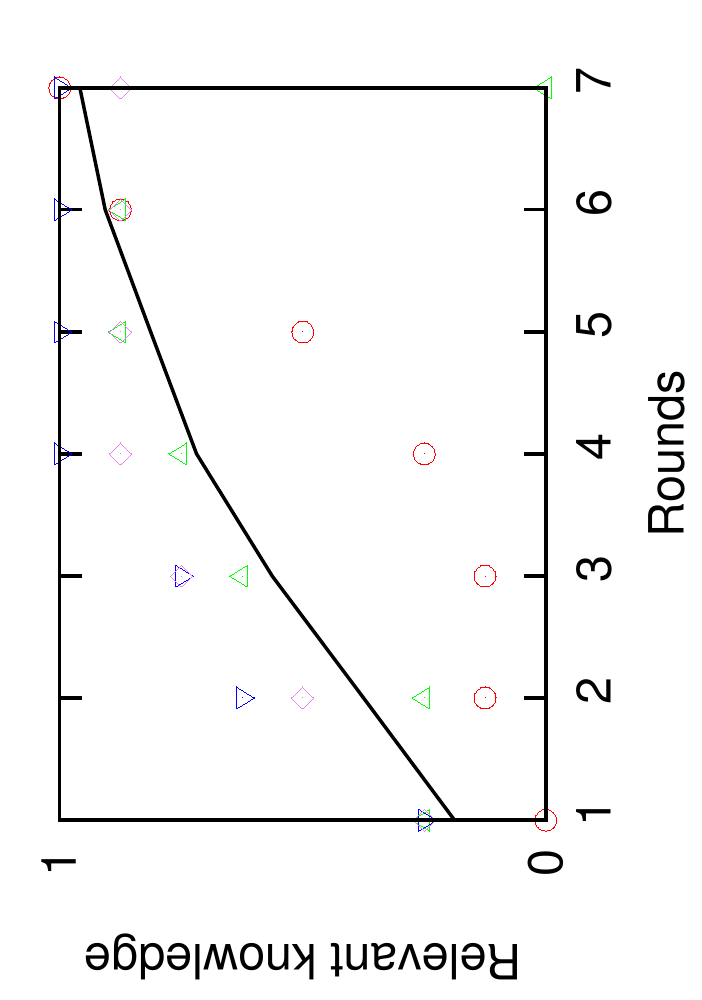}}

\put(40,170){\includegraphics[height=5.3cm,angle=-90,trim= 0cm 1.7cm 1.4cm 0pt,clip]{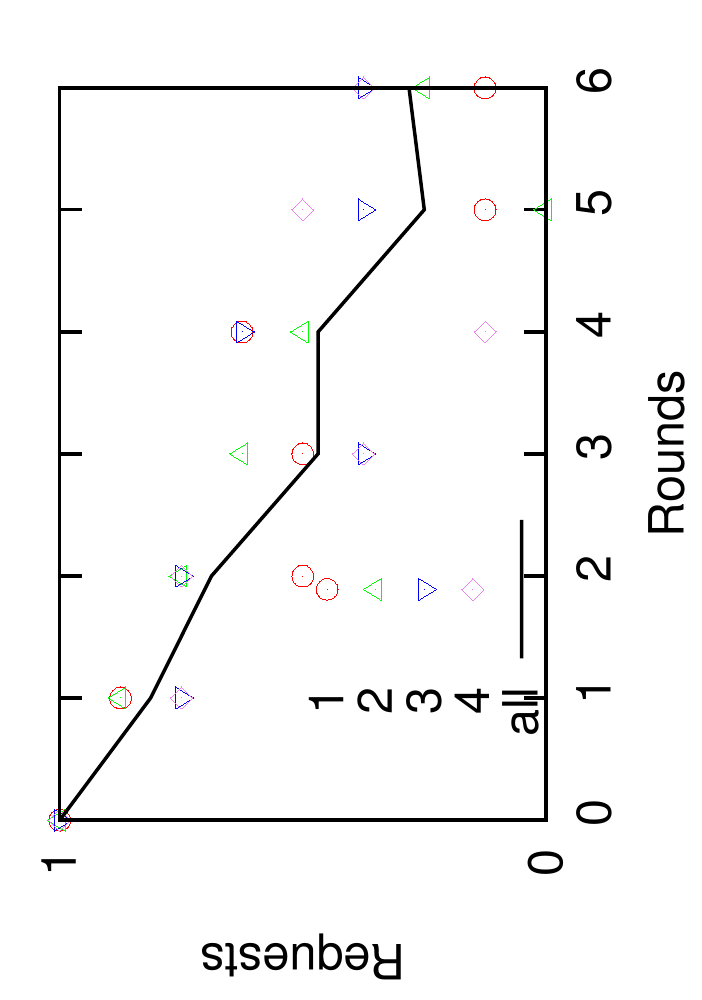}}
\put(40,100){\includegraphics[height=5.3cm,angle=-90,trim= 0cm 1.7cm 1.4cm 0pt,clip]{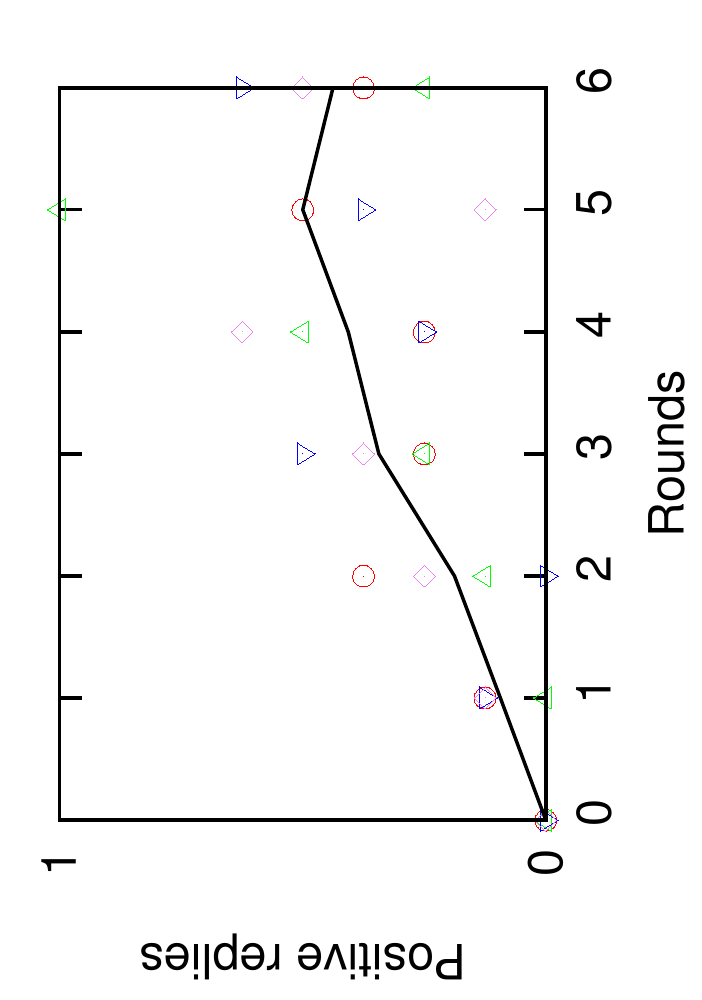}}
\put(40,30){\includegraphics[height=5.3cm,angle=-90,trim= 0cm 1.7cm 1.4cm 0pt,clip]{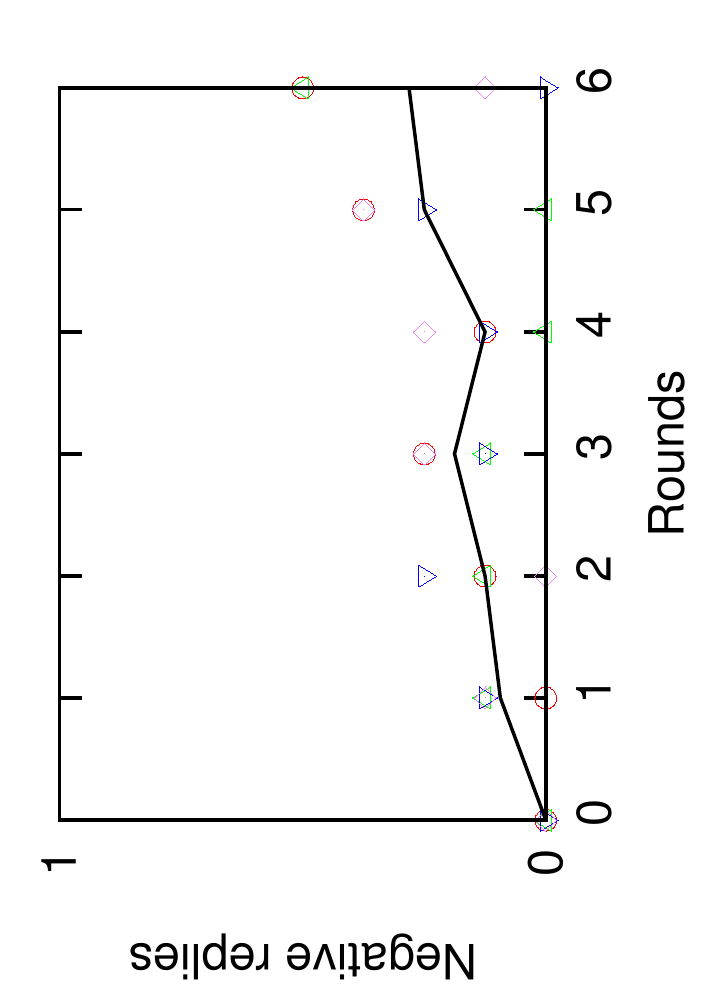}}
\put(40,-40){\includegraphics[height=5.3cm,angle=-90,trim= 0cm 1.7cm 1.4cm 0pt,clip]{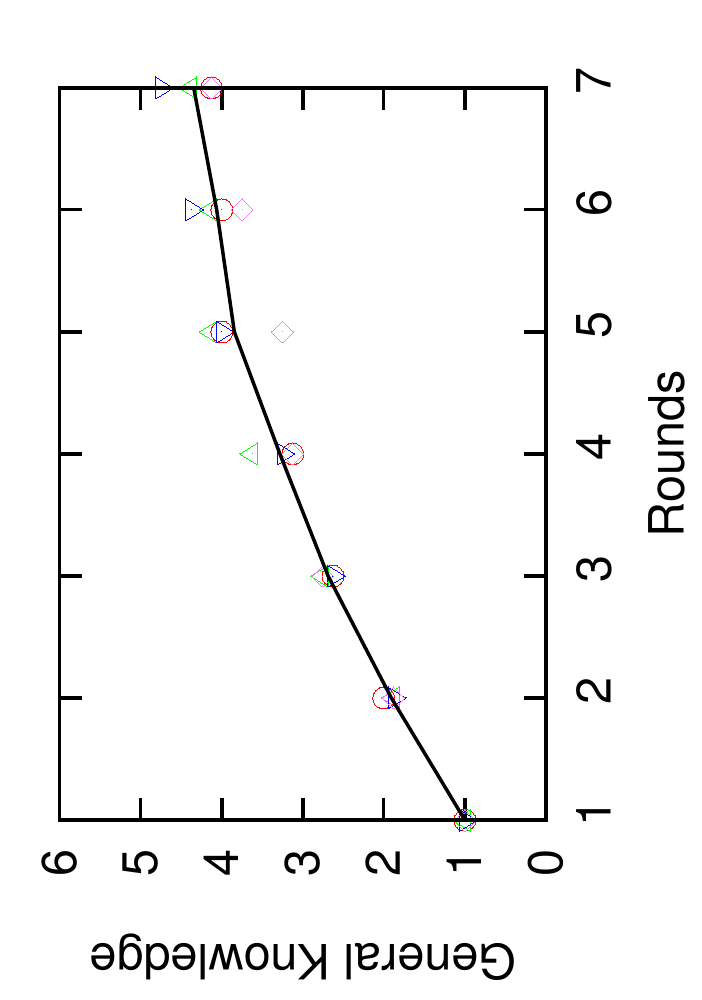}}
\put(40,-110){\includegraphics[height=5.3cm,angle=-90,trim= 0cm 1.7cm 0cm 0pt,clip]{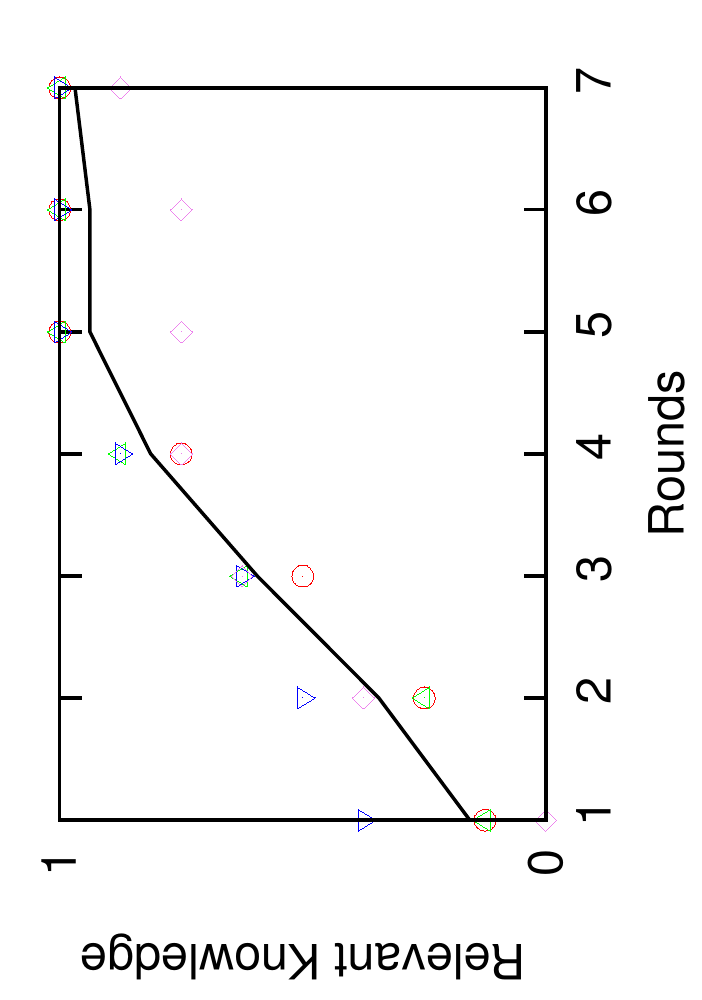}}

\put(-80,165){{\bf a}}
\put(40,165){{\bf b}}
\end{overpic}
\vspace{290pt}
\caption{{\bf Behavior as function of time step.}  
Number of messages per person for questions, positive (Y or R) and negative (N) answers for games 1 (red circles) and 2--4 (green, blue, purple symbols) and average of all games (black lines).
{\bf a}, Experimental data. {\bf b}, Modeled data.
}
\label{fig:function_of_timestep}
\end{center}
\end{figure*}
\end{center}

We also analyzed the participants' behavior as a function of round within each game (Fig.~\ref{fig:function_of_timestep}). 
During the first round, the only allowed message is a request, therefore all participants send messages during the first round.
The rate of requests then decays approximately linearly with time. 
As information about the participants' expertise is passed along with requests, information gained about others increases (Fig.~\ref{fig:function_of_timestep}d).
The rate of confirmations and referrals reaches comparatively low, but non-zero values after the second round (Fig.~\ref{fig:function_of_timestep}b), a similar behavior is observed for the negations (Fig.~\ref{fig:function_of_timestep}c).
Fig.~\ref{fig:function_of_timestep}e shows the time dependence of the fraction of relevant knowledge that was received, i.e. knowledge of the desired expert. 
Interestingly, almost all players learn the identity of the expert needed to complete their task after $4$ to $6$ rounds in all games but the first. 
This is remarkable, because the value of the overall knowledge gained about others reaches only approximately $0.5$ during the same time (Fig.~\ref{fig:function_of_timestep}d).
This means, that indeed the communication network is capable of channeling desired information more quickly to its recipient than by random messages. 
The network could be said to have acquired the ability of efficient information retrieval.

{\bf Modeling results ---} We simulated games with the Bayesian agents described by our model. The modeled agents were homogeneous regarding their personality traits and --- to yield agreement with the experimental data --- had to be given strong cooperative preference for referrals and confirmations ($\alpha=1$ , $\beta=5$ and $\gamma=1$). 
The network formed by the simulated players is shown in Fig.~\ref{fig:interaction_networks_model}. 
As in the experimental data (Fig.~\ref{fig:interaction_networks}), a heterogeneous communication network forms over the four games with generally reciprocal communication links between agents.
The comparison of the correlation matrix (Fig.~\ref{fig:correlation_matrix}) again shows reasonable agreement with the experimental data, however the temporal correlation between games is not as pronounced as in the experiment.

Within the noise resulting from the small number of games played, the agreement regarding the dependence on timestep shown in Fig.~\ref{fig:function_of_timestep} is remarkably good.

\section*{Conclusions}
Our experiment shows that cooperation arises in a setting where subjects communicate only through an electronic interface and no face-to-face contact or emotional content of messages are possible.
We find that stable channels of communication form between subjects that are robust against repeated perturbations of their network. 
These channels can be viewed as robust treaties between the partners, with replies issued signaling the willingness to cooperate, but also the expectation of receiving help, in the future.

This interpretation is supported by the pronounced reciprocity in the network of the stable links.
Along the links, specific information is acquired by individuals more efficiently than random information --- the established network acts as an efficient information retrieval system.
Once stable links are acquired, maintaining these becomes of high priority for individuals:
Subjects are willing to sacrifice substantial time-resources in order to reply to requests with referrals, confirmations and even negations. 
Especially negations carry little direct information content but do purvey an indication of the willingness to cooperate in the future. 
In that sense, these messages may be a way of re-enforcing existing collaborations. 

In conclusion, frequently and efficiently communicating with some communication partners necessarily leads to the neglect of others. 
Our results suggest that human communication is by no means a generous process and that cooperation does not emerge automatically, but requires investments in the establishment of new partnerships and expenses for the maintenance of established ties.

\begin{materials}
\section{Experimental data} The social experiment was carried out at the Center for Models of Life, Niels Bohr Institute Copenhagen, Denmark, in October 2013. 
Subjects were researchers with a background in theoretical physics. 
Identities were anonymized.
The game was played with four repetitions during a period of approximately 2 hours. 
\section{Computer Simulations} All data analysis and computer simulations were carried out on standard desktop computers.
\end{materials}

\begin{acknowledgments}
The authors acknowledge financial support by the Danish National Research Foundation through the Center for Models of Life.
\end{acknowledgments}


\end{article}

\begin{center}
\begin{figure*}[t]
\begin{center}
\begin{overpic}[width=5cm,angle=-90,trim= 0cm 0pt 0pt 0pt,clip]{dummy.pdf}
\put(-120,0){\includegraphics[width=4cm,angle=0]{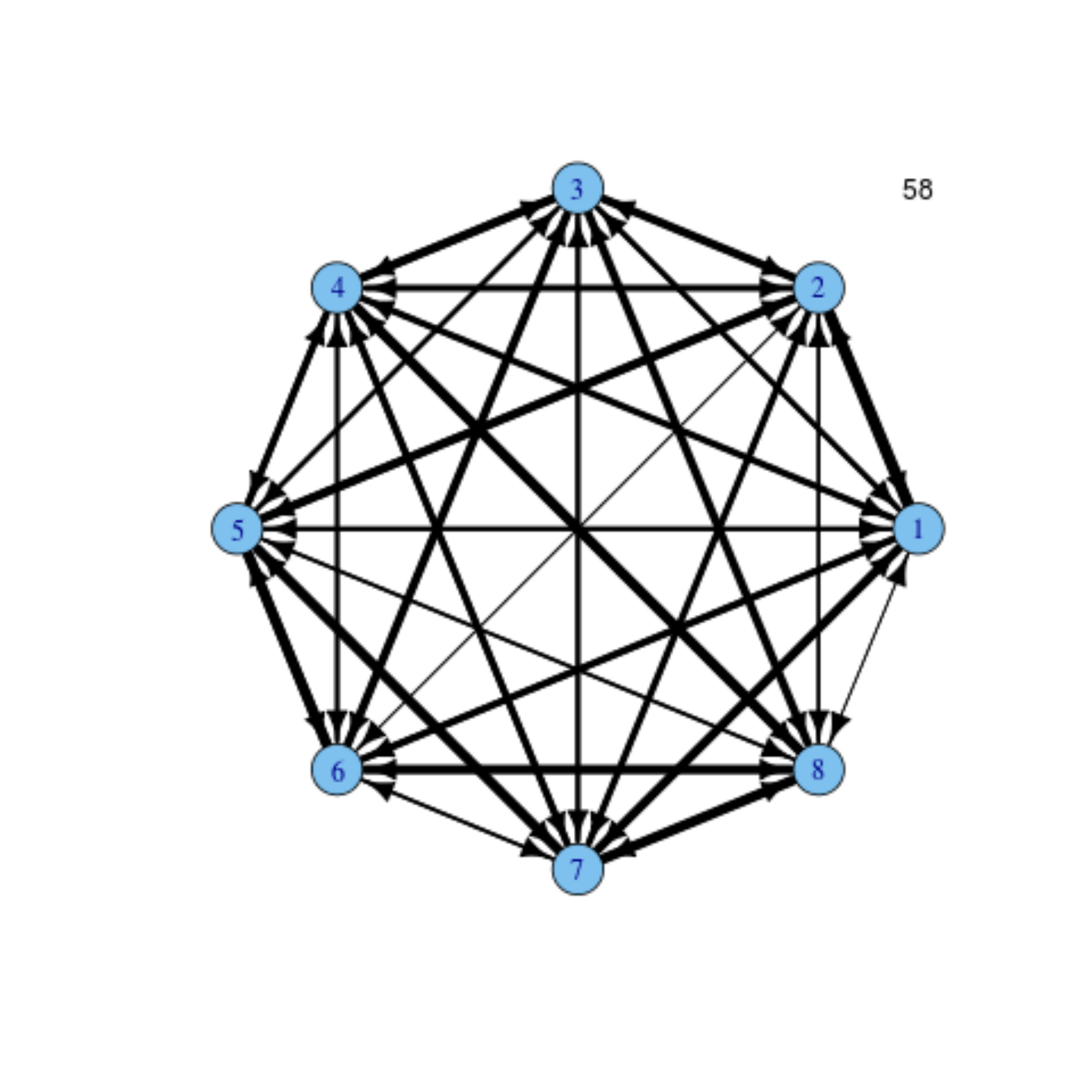}}
\put( -50,0){\includegraphics[width=4cm,angle=0]{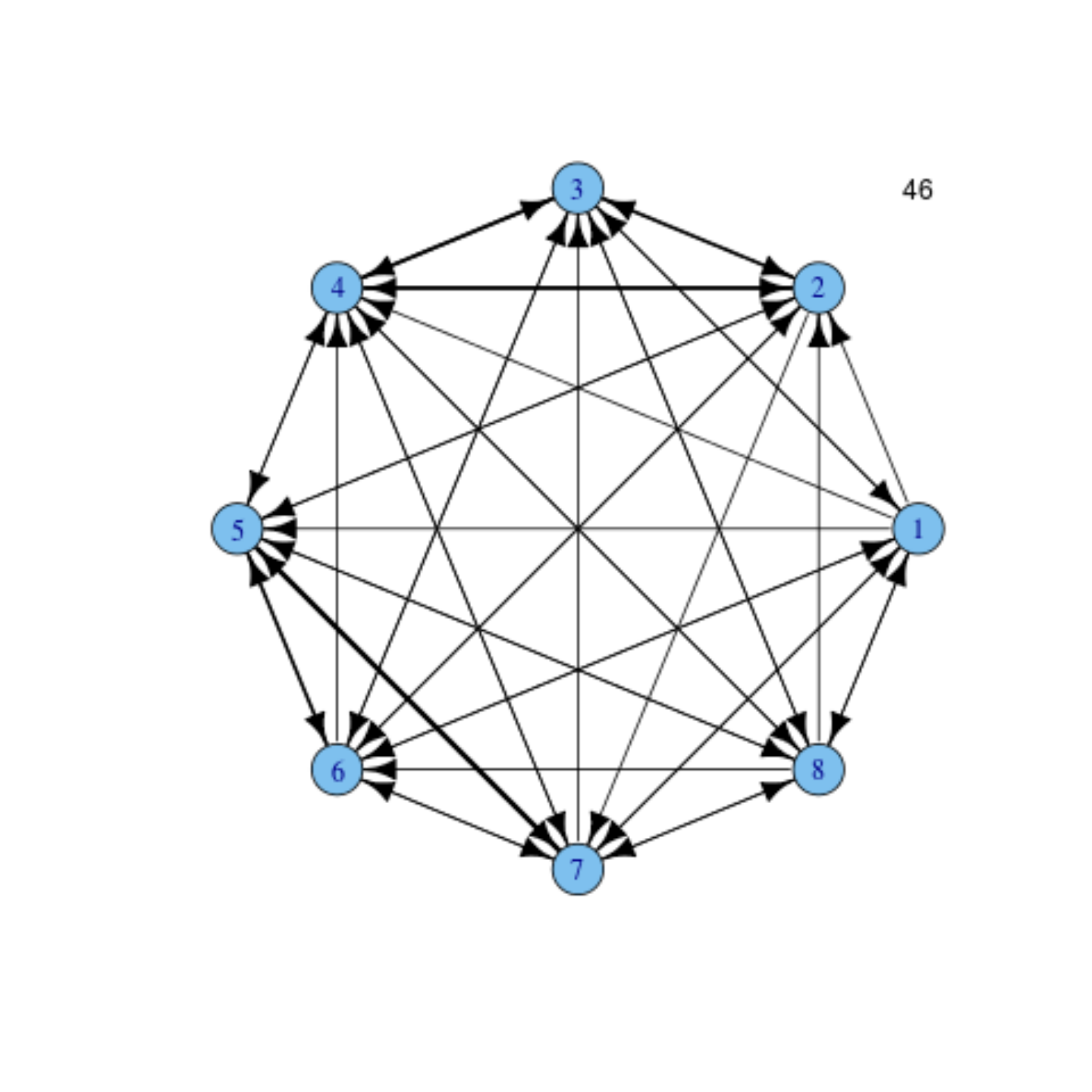}}
\put(  20,0){\includegraphics[width=4cm,angle=0]{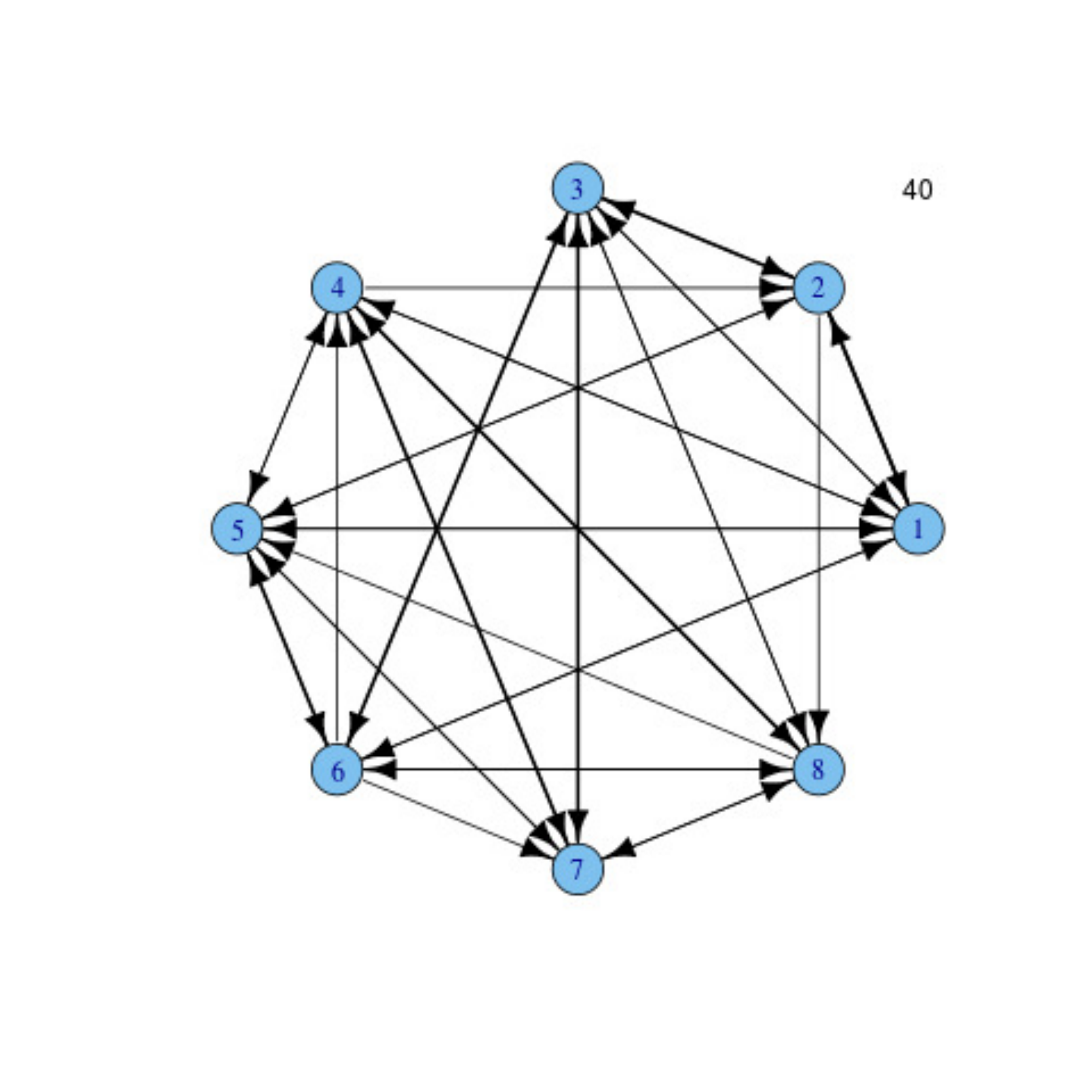}}
\put(  90,0){\includegraphics[width=4cm,angle=0]{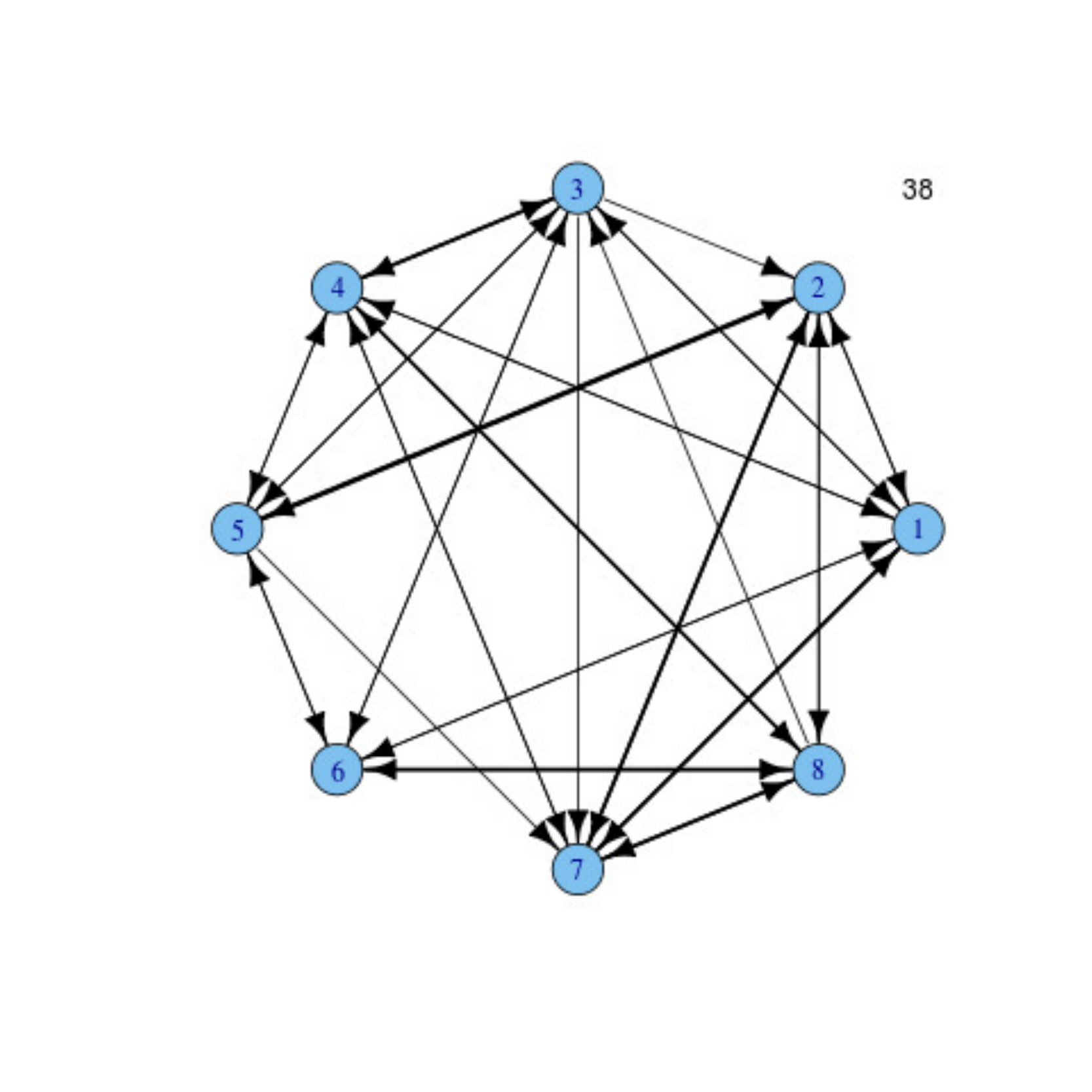}}
\put(  160,0){\includegraphics[width=4cm,angle=0]{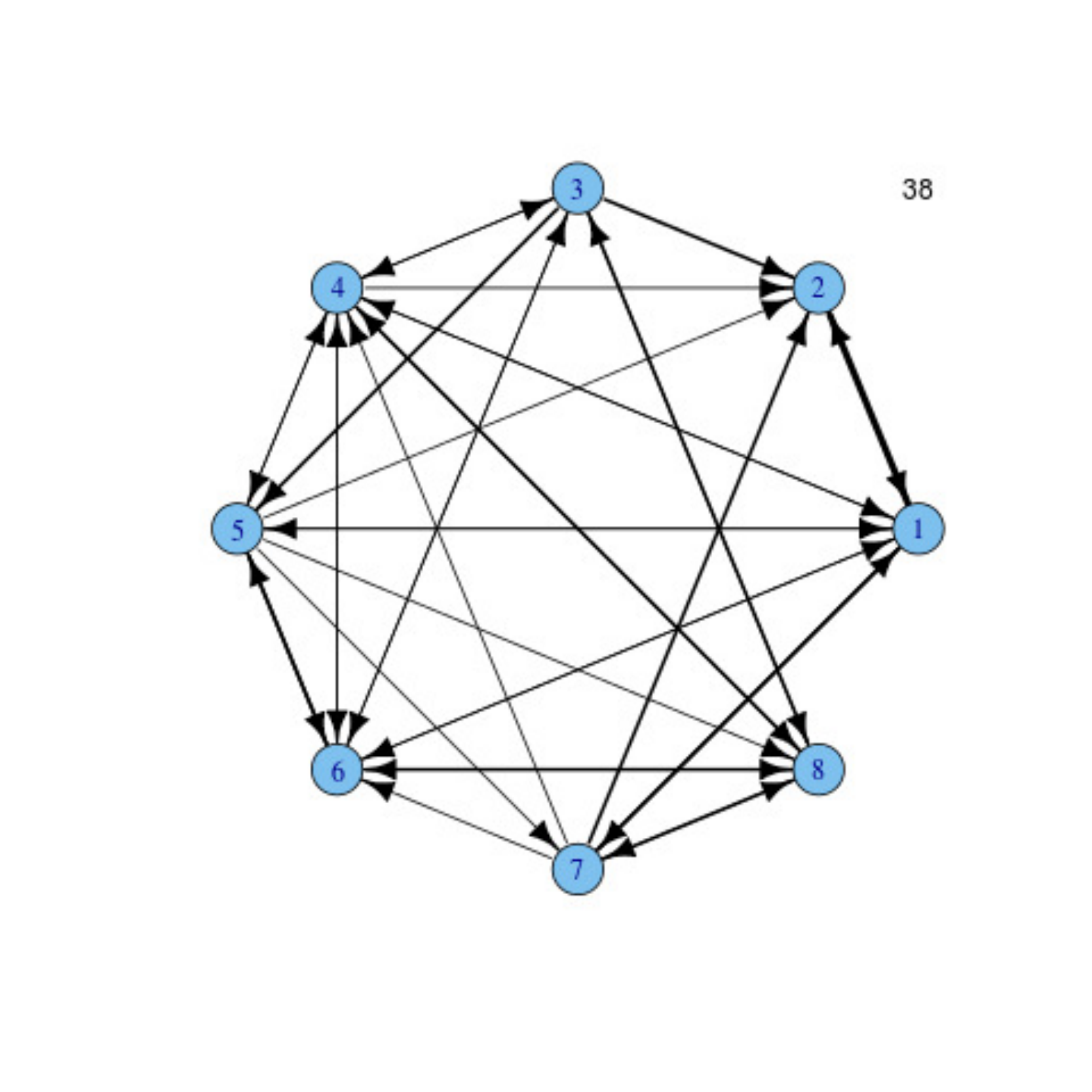}}

\put(-120,0){\includegraphics[width=4.4cm,angle=-90]{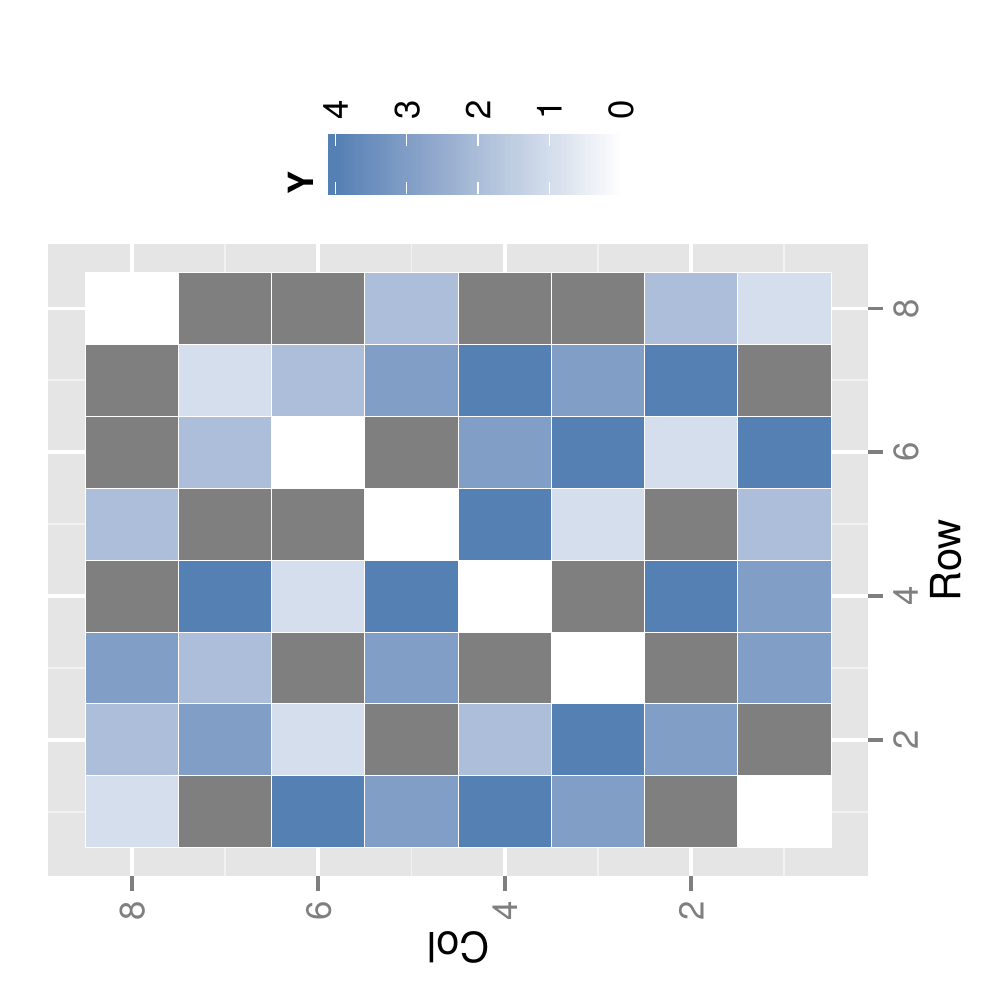}}
\put( -50,0){\includegraphics[width=4.4cm,angle=-90]{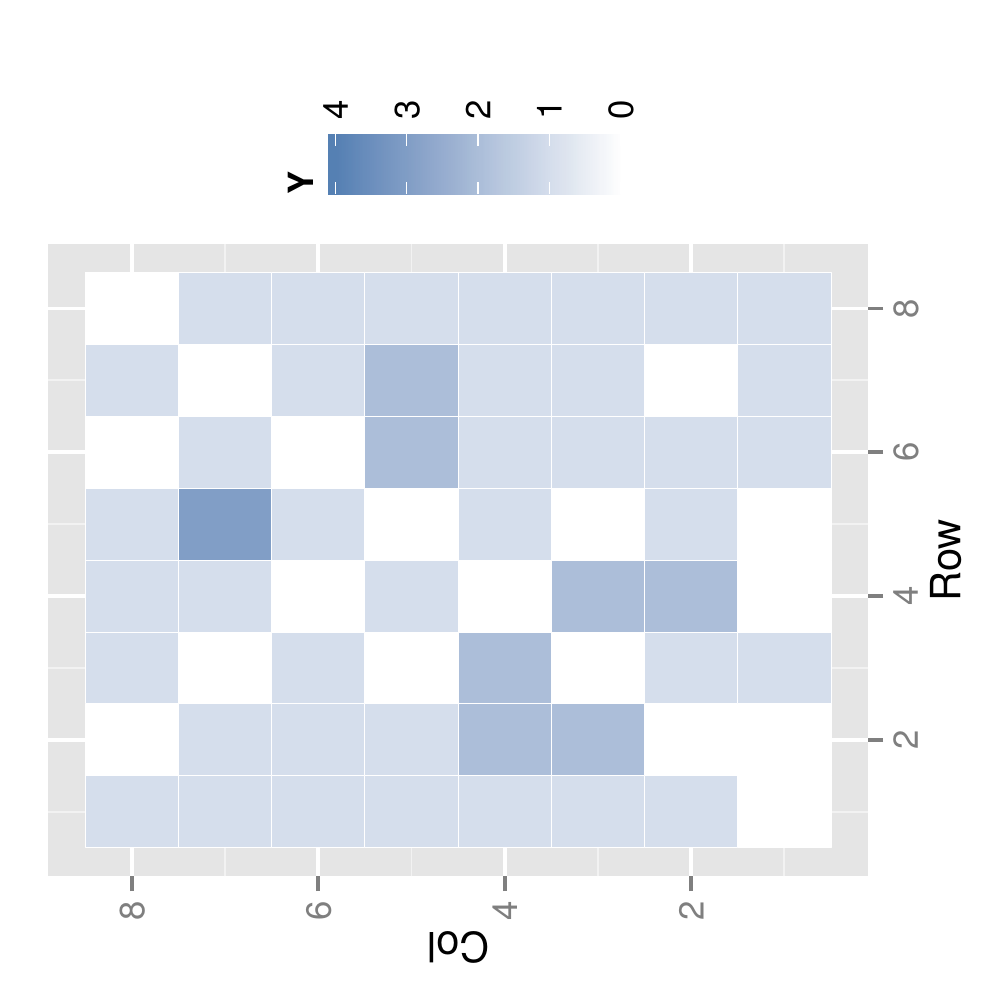}}
\put(  20,0){\includegraphics[width=4.4cm,angle=-90]{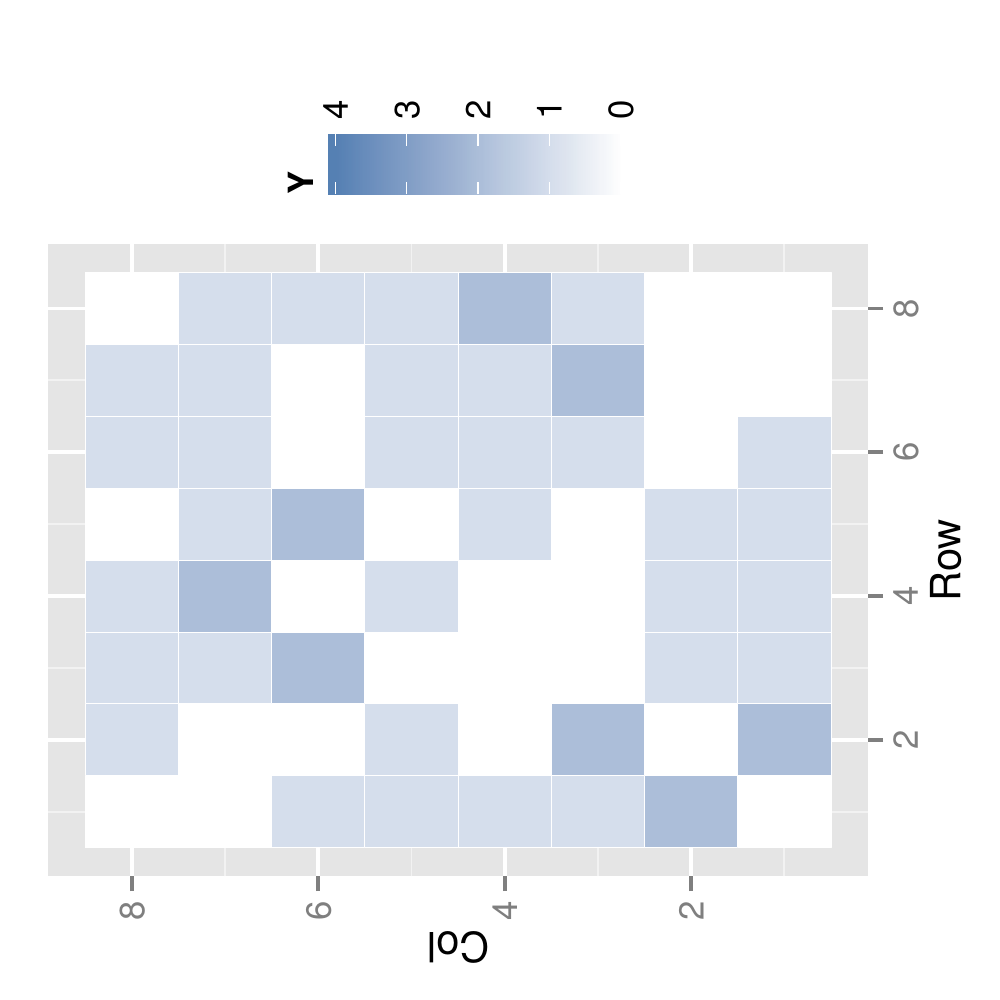}}
\put(  90,0){\includegraphics[width=4.4cm,angle=-90]{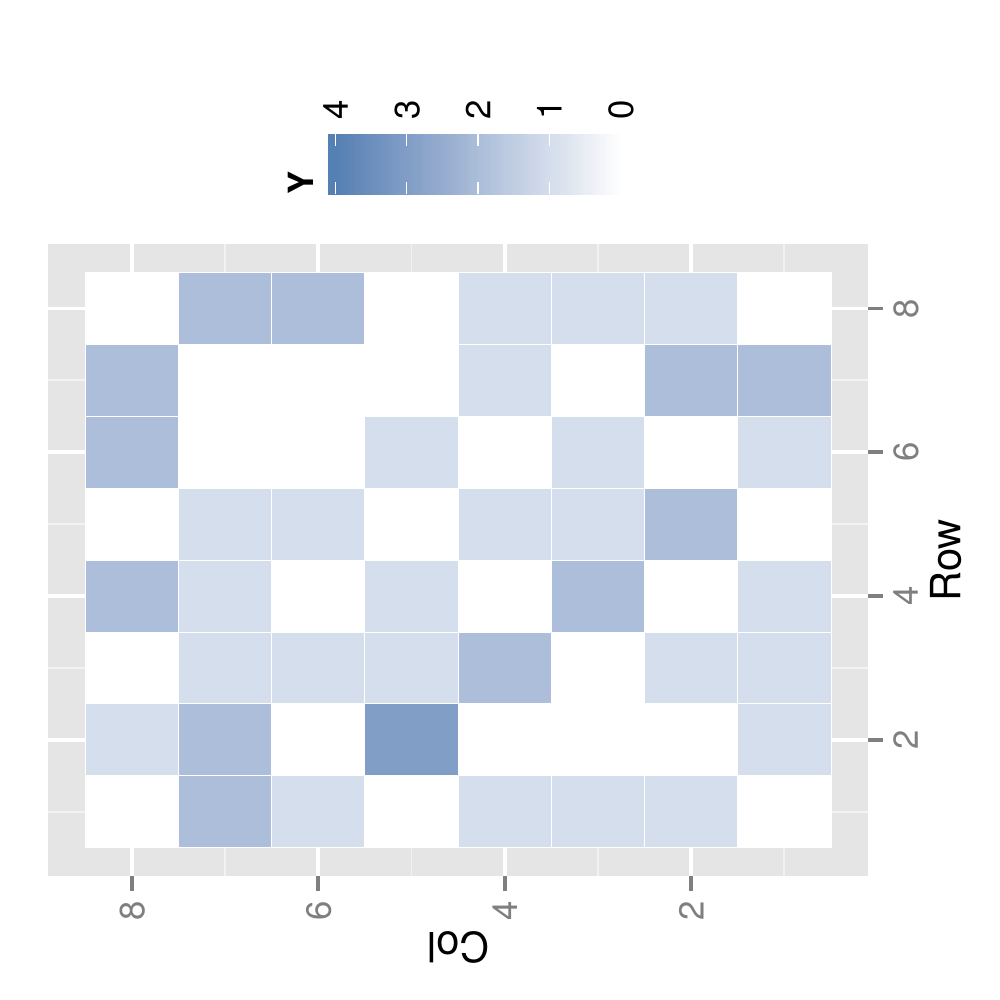}}
\put(  160,0){\includegraphics[width=4.4cm,angle=-90]{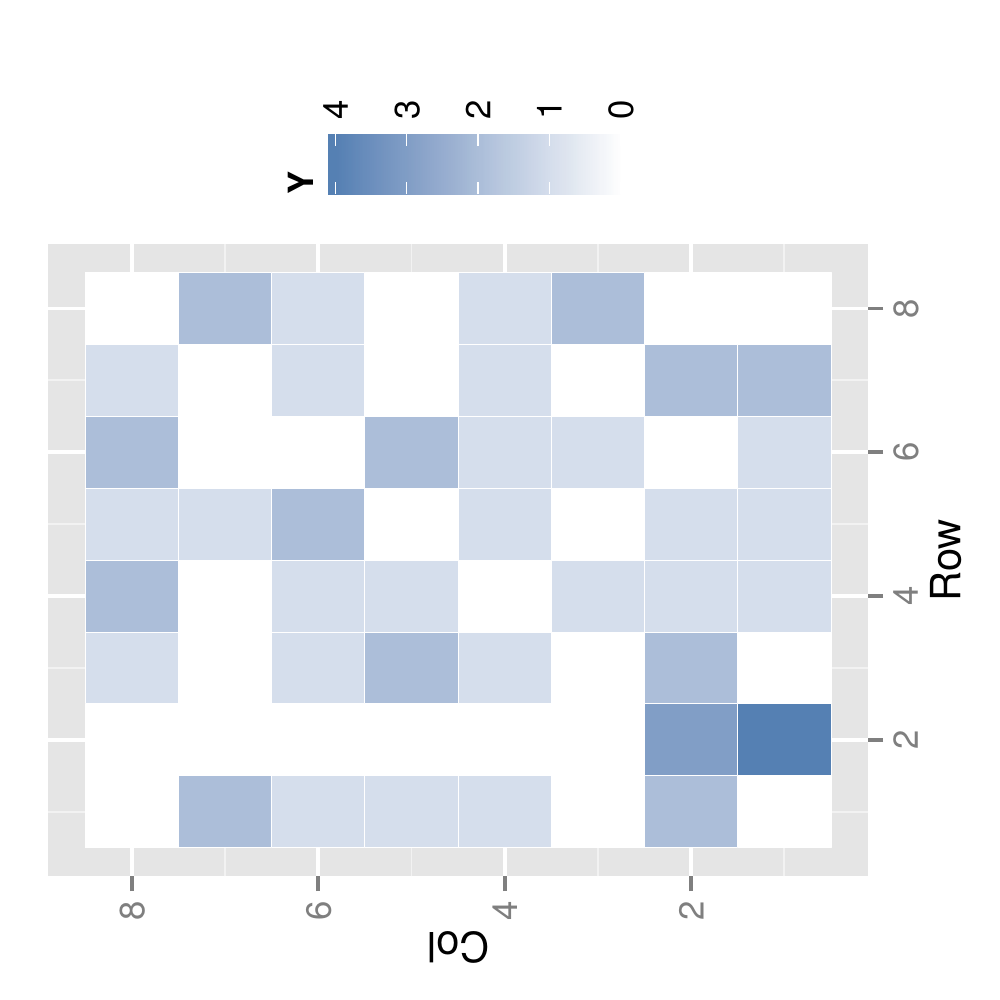}}

\put(-120,-100){\includegraphics[width=4.4cm,angle=-90]{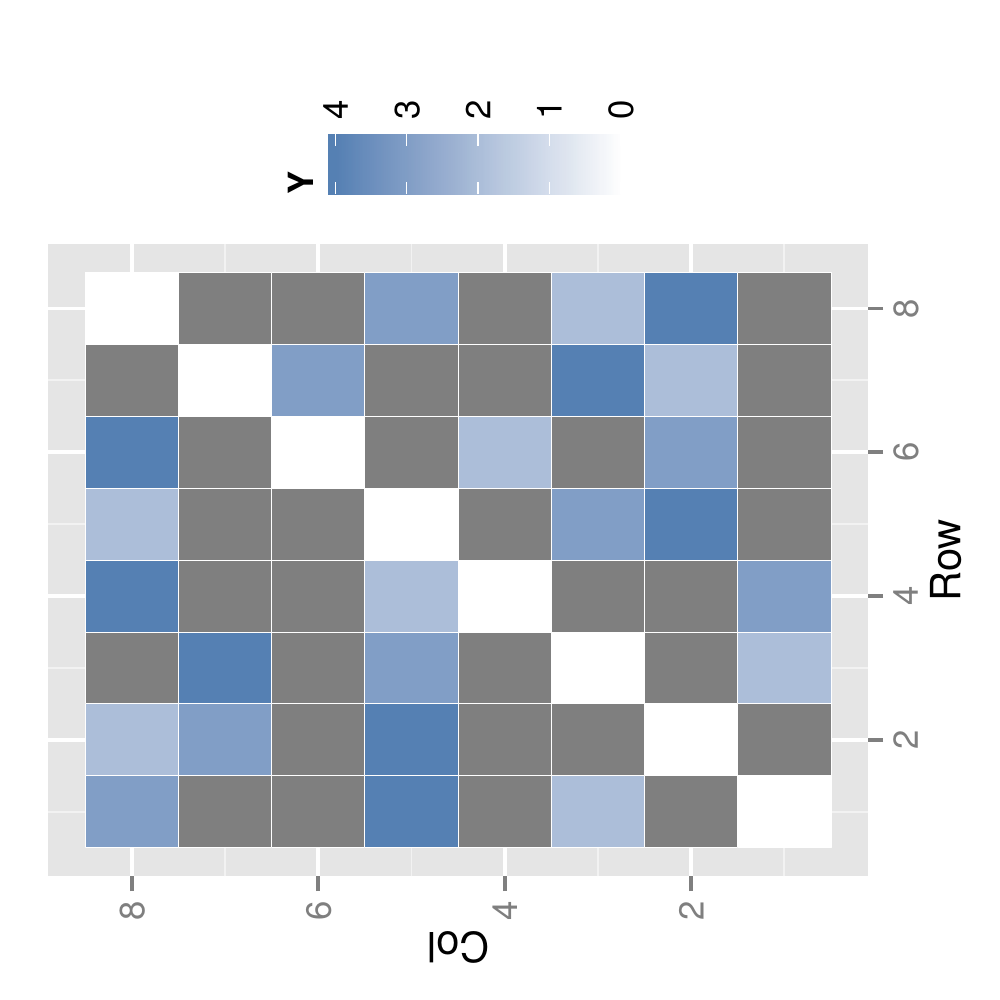}}
\put(-110,80){{\bf a}}
\put(-40,80){{\bf b}}
\put(30,80){{\bf c}}
\put(100,80){{\bf d}}
\put(170,80){{\bf e}}
\put(-110,-100){{\bf f}}

\end{overpic}
\vspace{260pt}
\caption{{\bf Experimental communication networks.}  
Adjacency matrices and corresponding networks for the entire duration (a) and games 1---4 separately (b---e).
Indices on axes denote players 1---8.
{\bf f}, Adjacency matrix resulting from the players' self-assessment data.
}
\label{fig:interaction_networks}
\end{center}
\end{figure*}
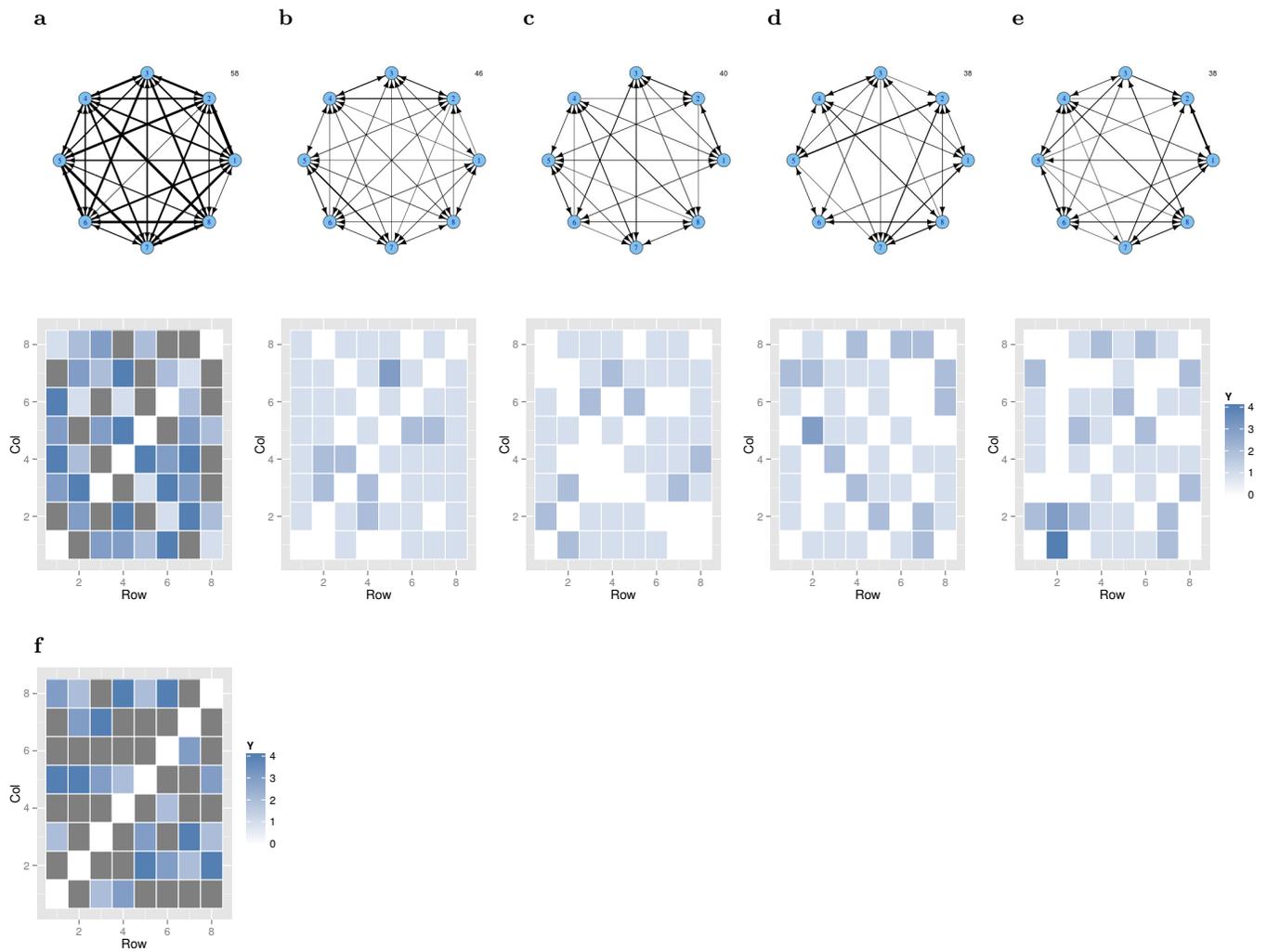
\end{center}

\begin{center}
\begin{figure*}[t]
\begin{center}
\begin{overpic}[width=5cm,angle=-90,trim= 0cm 0pt 0pt 0pt,clip]{dummy.pdf}
\put(-120,0){\includegraphics[width=4cm,angle=0]{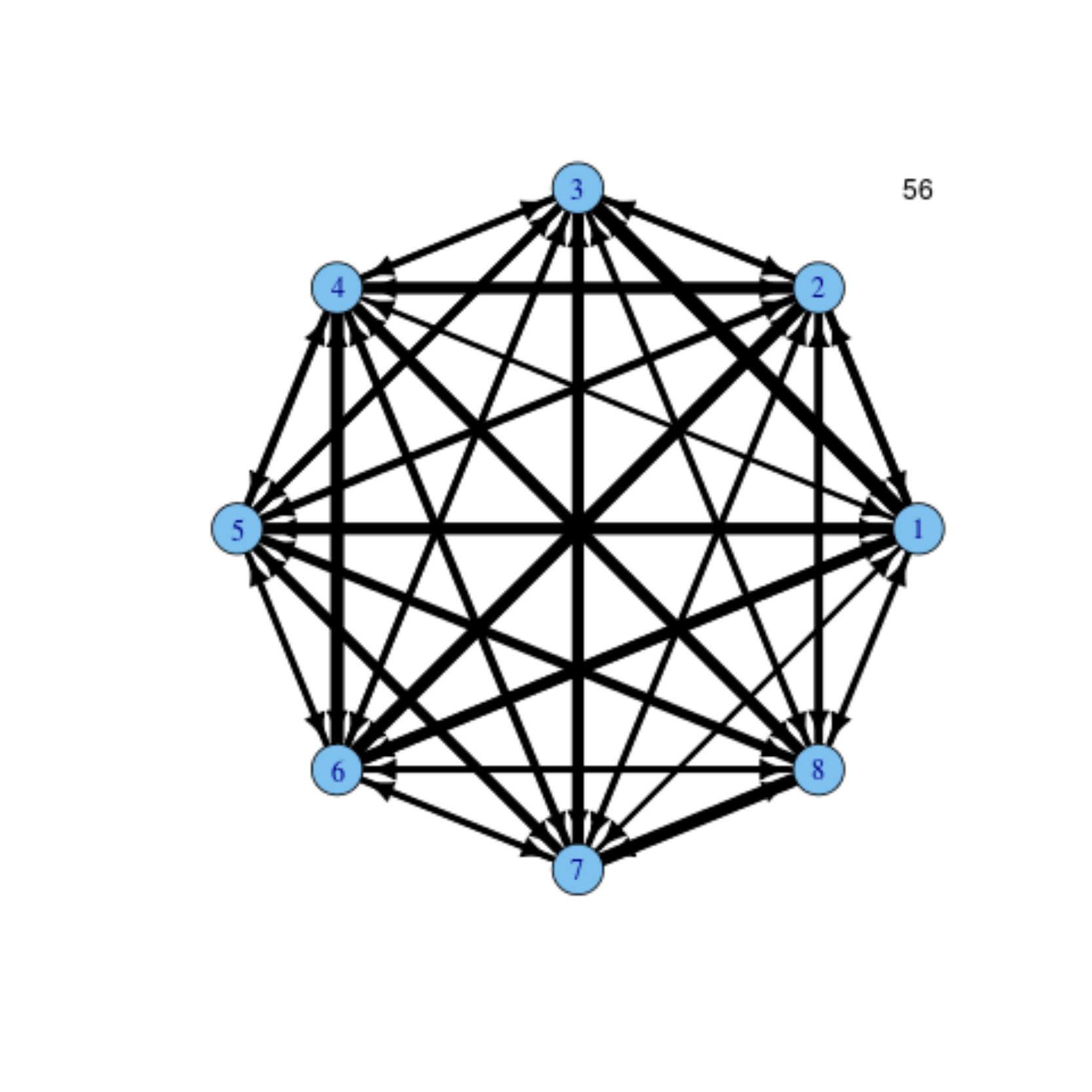}}
\put( -50,0){\includegraphics[width=4cm,angle=0]{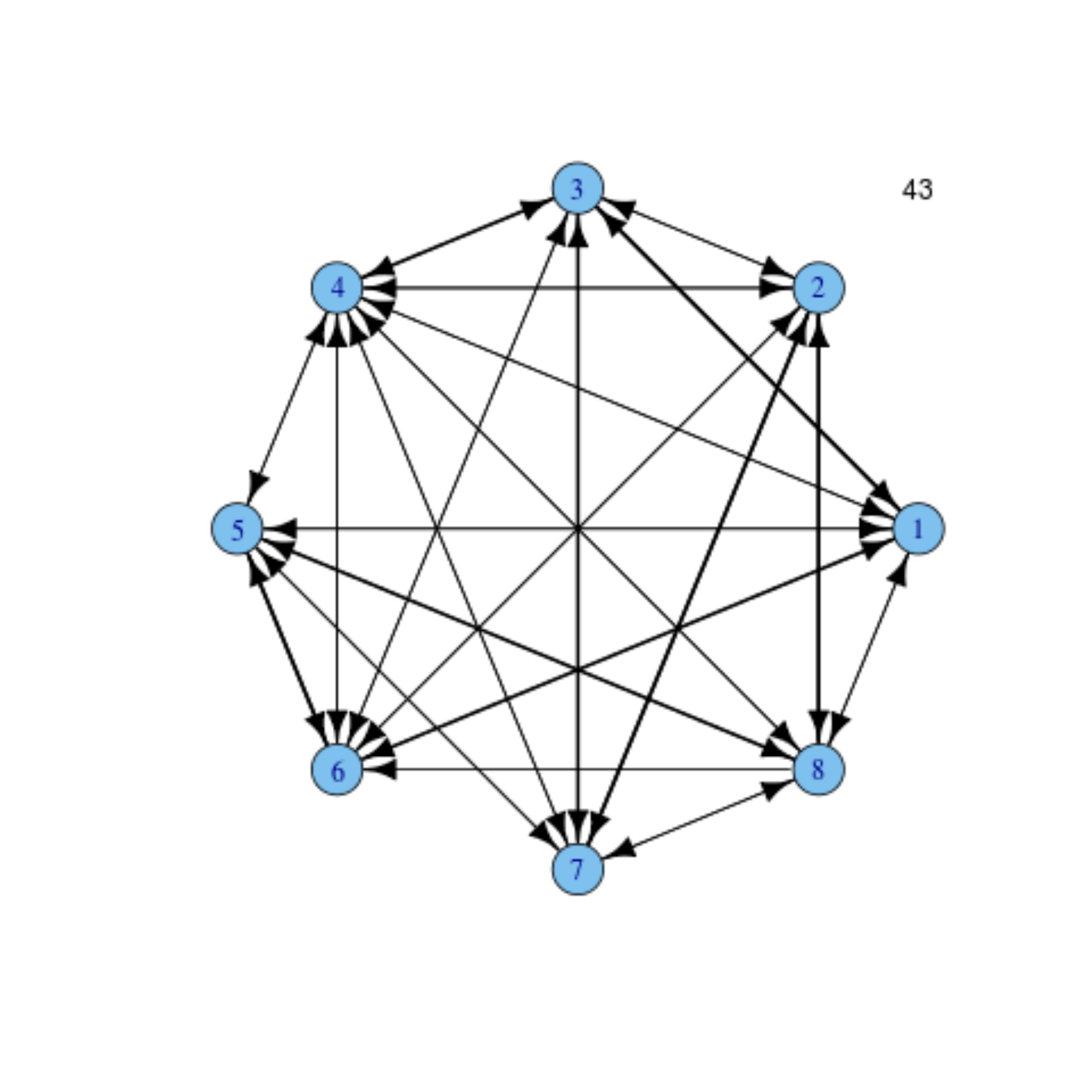}}
\put(  20,0){\includegraphics[width=4cm,angle=0]{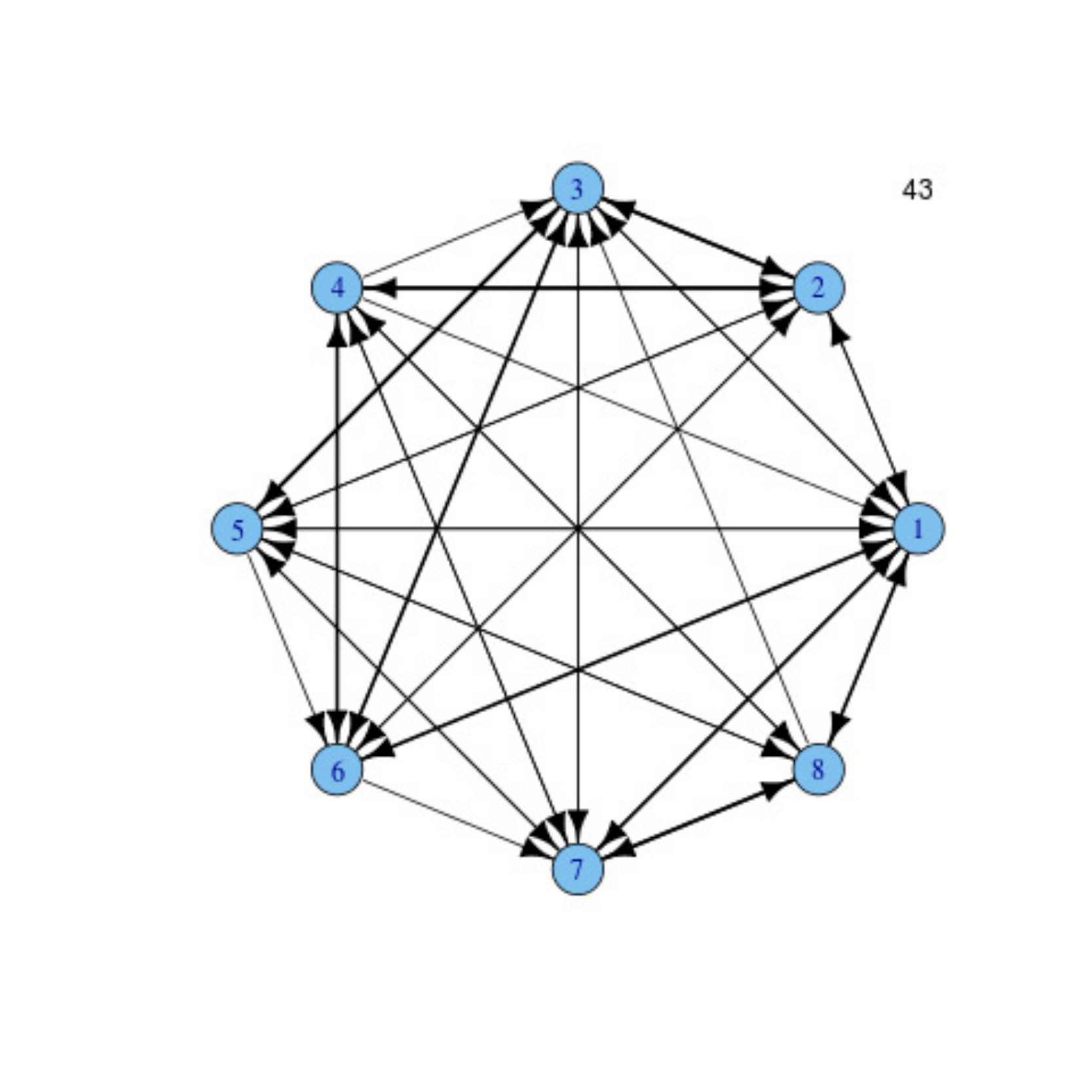}}
\put(  90,0){\includegraphics[width=4cm,angle=0]{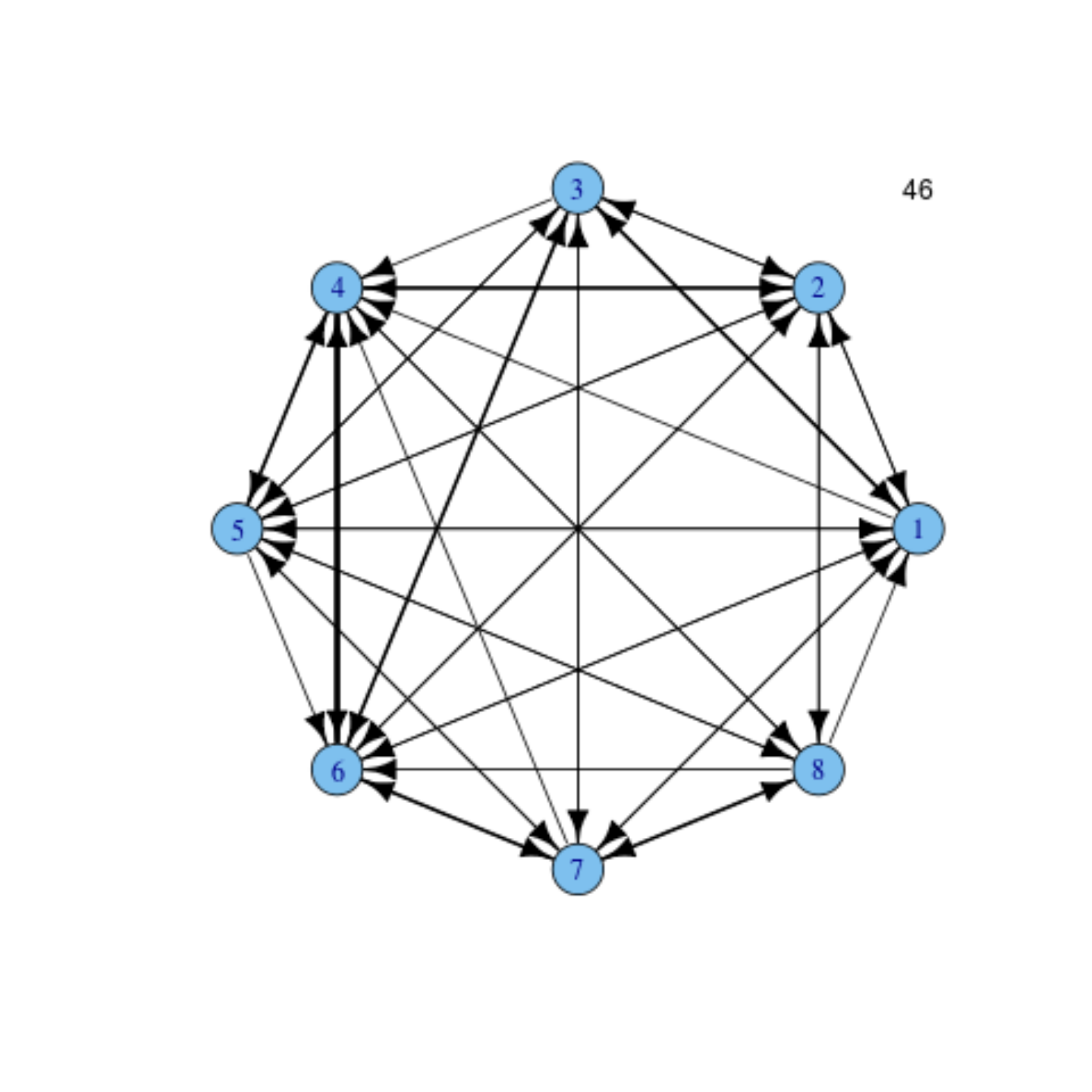}}
\put(  160,0){\includegraphics[width=4cm,angle=0]{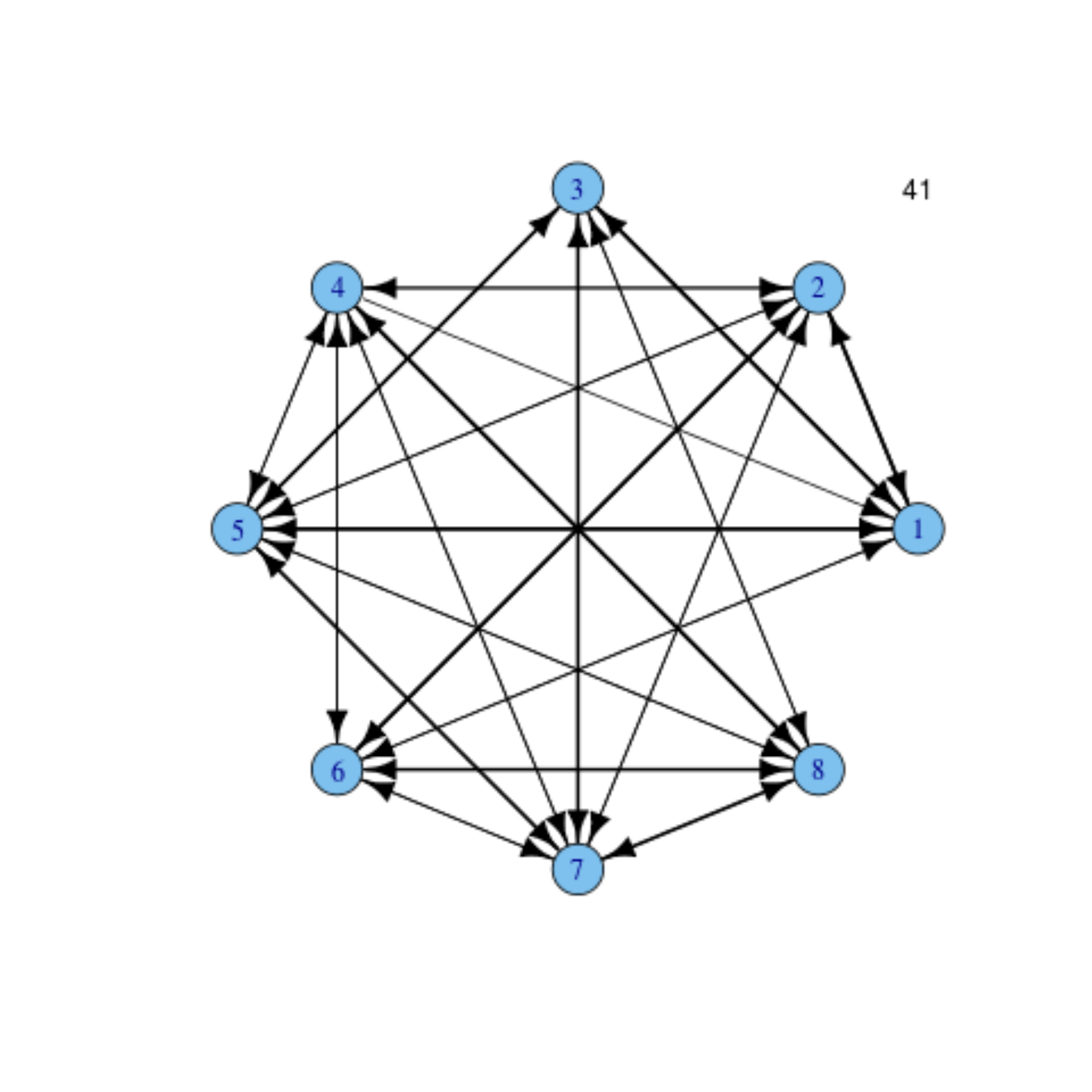}}

\put(-120,0){\includegraphics[width=4.4cm,angle=-90]{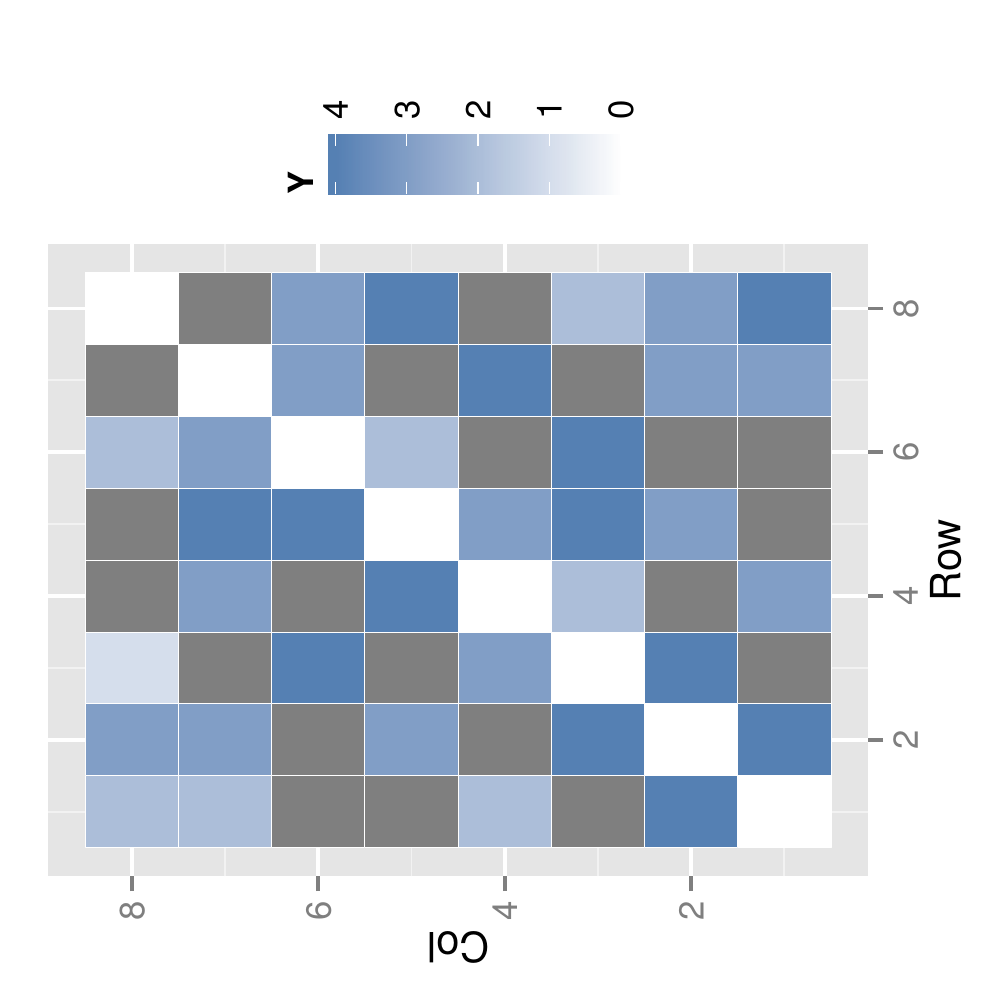}}
\put( -50,0){\includegraphics[width=4.4cm,angle=-90]{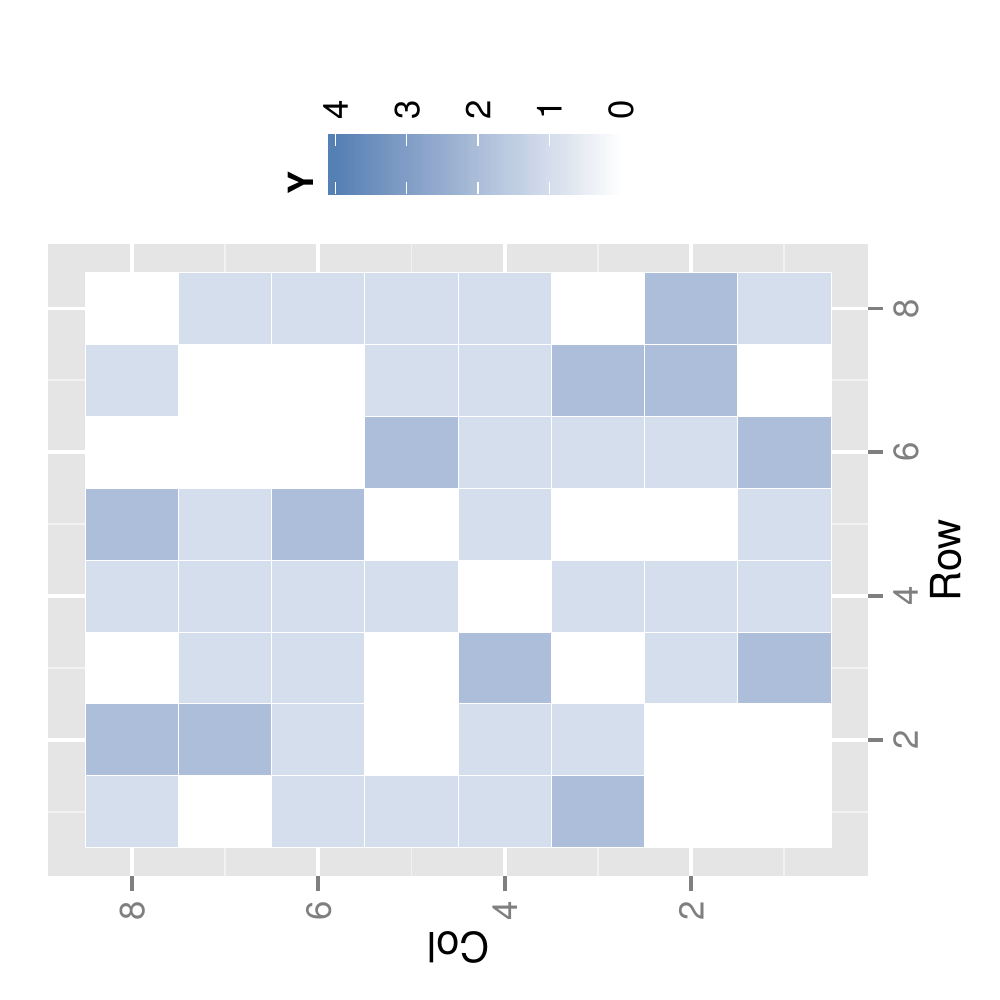}}
\put(  20,0){\includegraphics[width=4.4cm,angle=-90]{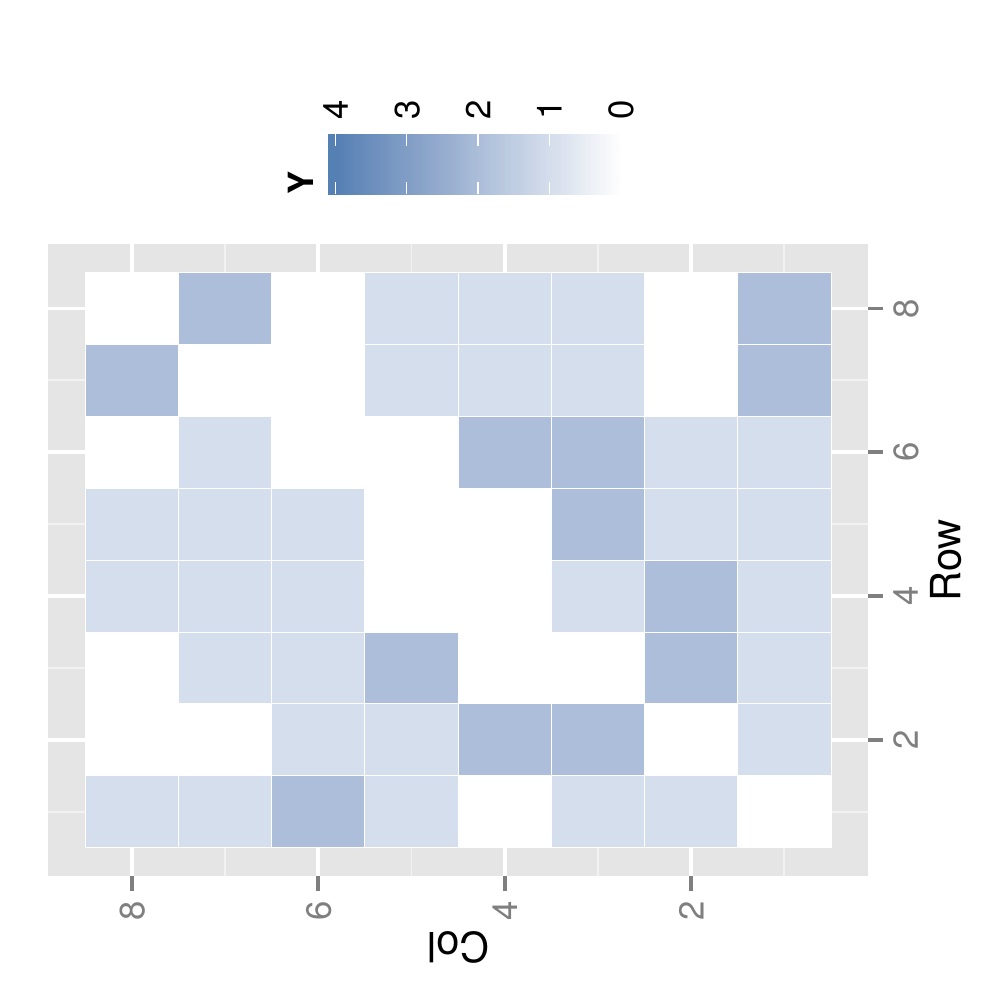}}
\put(  90,0){\includegraphics[width=4.4cm,angle=-90]{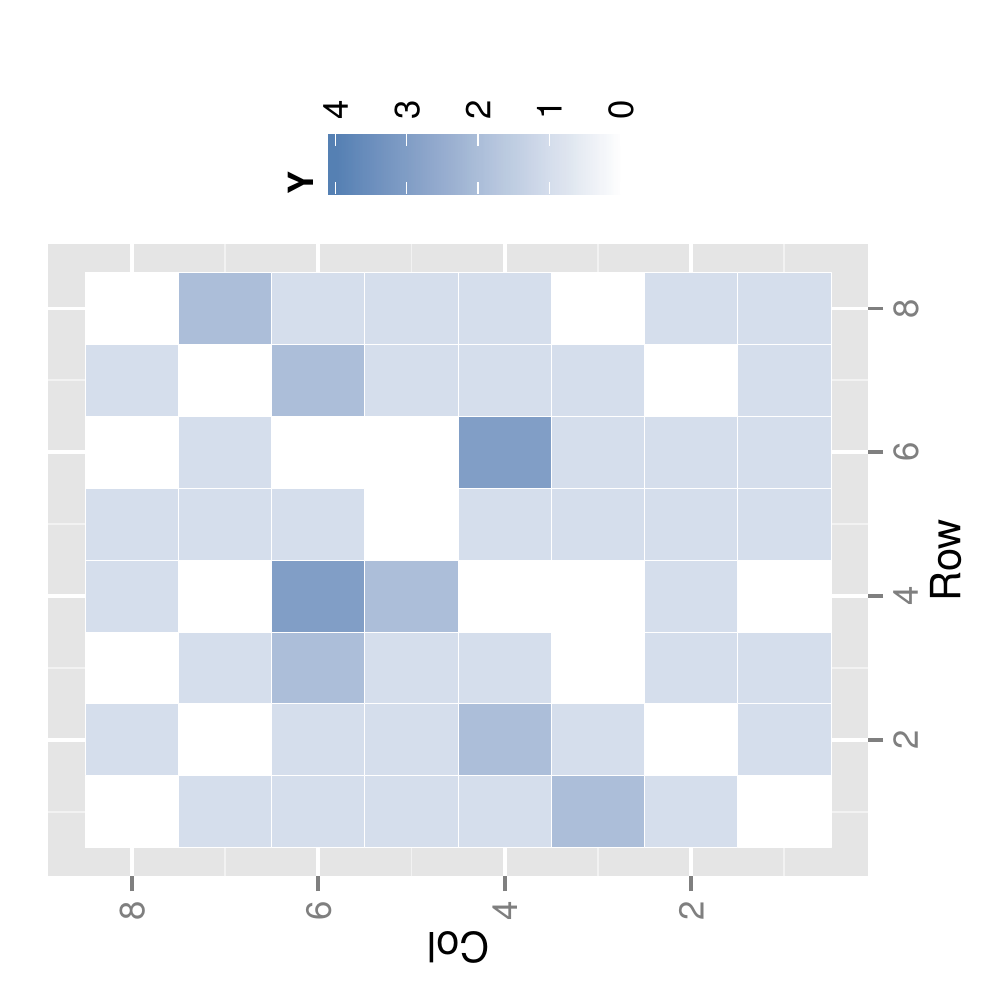}}
\put(  160,0){\includegraphics[width=4.4cm,angle=-90]{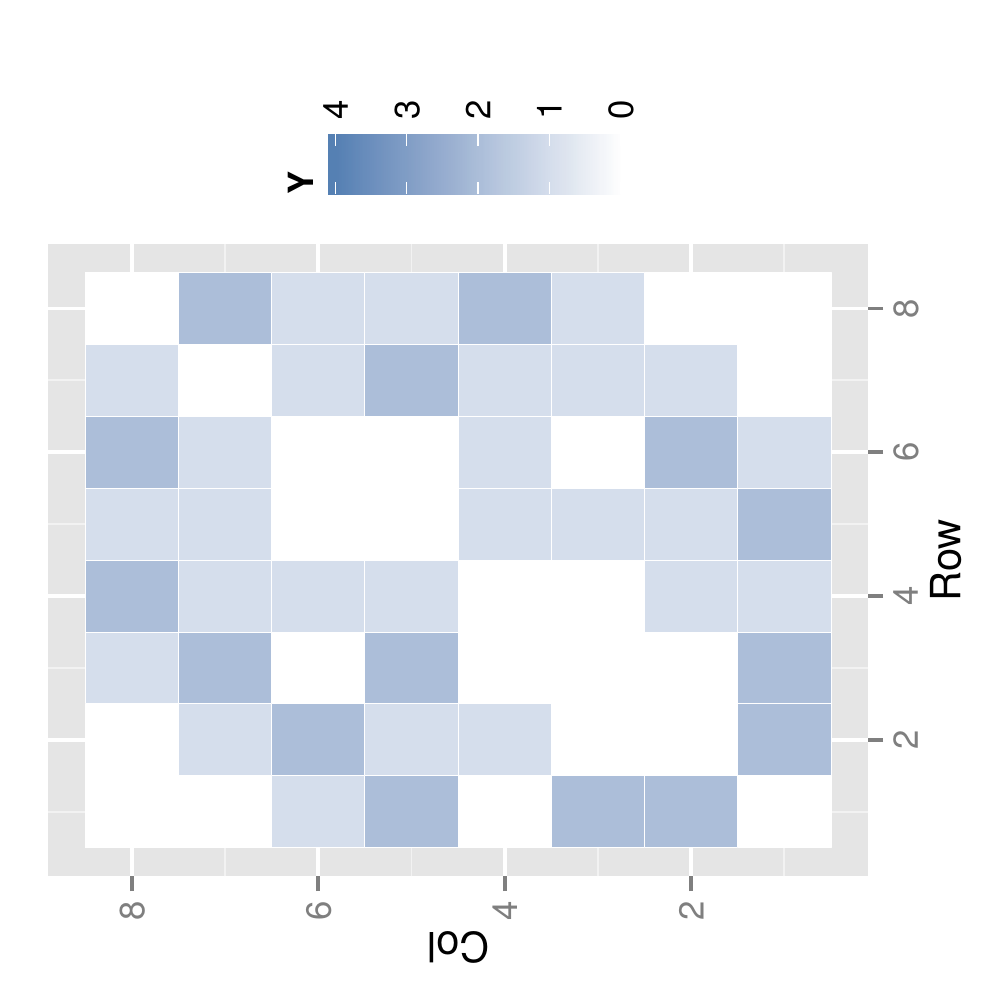}}

\put(-110,80){{\bf a}}
\put(-40,80){{\bf b}}
\put(30,80){{\bf c}}
\put(100,80){{\bf d}}
\put(170,80){{\bf e}}

\end{overpic}
\vspace{130pt}
\caption{{\bf Model communication networks.}  
Adjacency matrices and corresponding networks for the entire duration (a) and games 1---4 separately (b---e).
Indices on axes denote players 1---8.
}
\label{fig:interaction_networks_model}
\end{center}
\end{figure*}
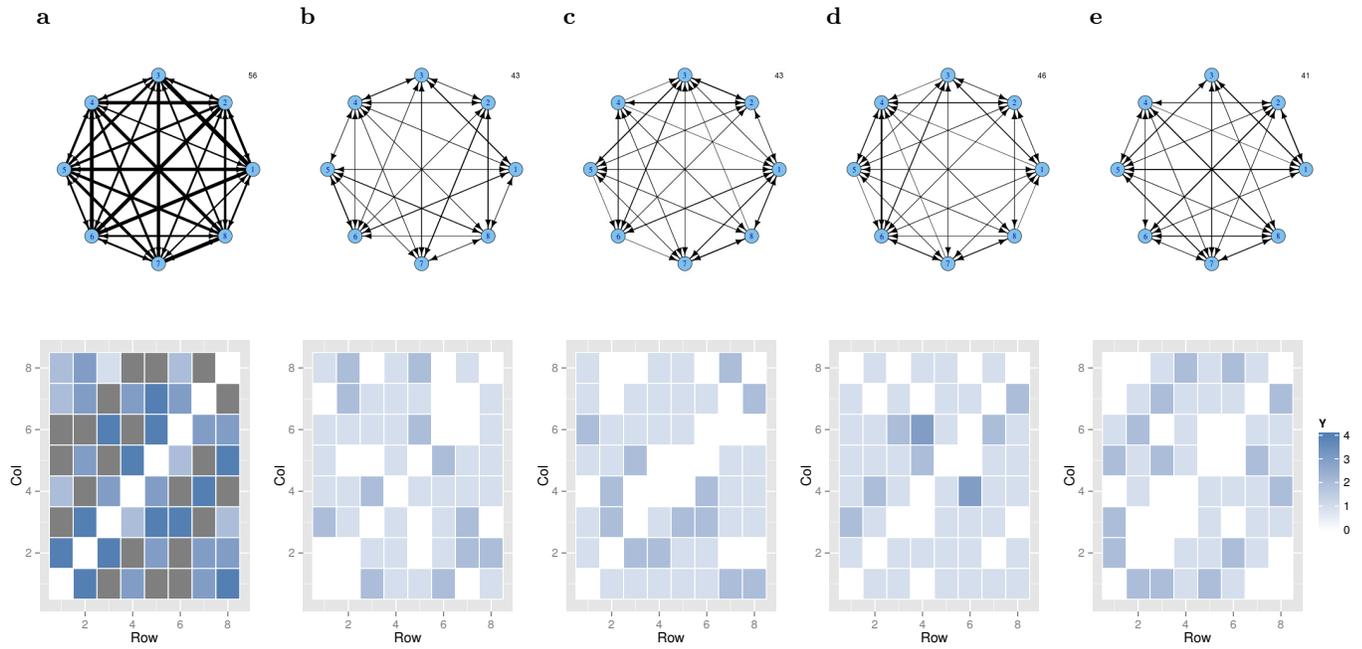
\end{center}


\begin{table}[h]
\begin{tabular}{|c|c|c|c|c|}
\hline
{\bf reply type} & {\bf average lag} & {\bf average rate} & {\bf average lag (Mod)} & {\bf average rate (Mod)}  \\
\hline
N & 2.50 & 0.24 & 2.64 & 0.25\\
\hline
Y & 2.28 & 0.33 & 2.73 & 0.46\\
\hline
no reply with knowledge & -- & 0.55  & -- & 0.65\\
\hline
\end{tabular}
\caption{{\bf Reply rates and times.} Reply Type, average lag in time steps conditional on reply and average rate of reply. Reply types are negative (N), positive (Y) and lack of reply at the end of a given round. For the latter, we give the fraction of occasions when the answer was known but not given.}
\label{tab:summary_table}
\end{table}


\appendix
\section{Choice of message}

If a modeled participant knows who his expert is, but has not sent him a request yet, he will always do so in the next round. If a participant knows his expert his preference to send requests $\alpha$ is also reduced to $0.1\%$ of its original value, until the next game.
Otherwise $P(M=(T,x,y),k)$ is the probability of a message of type $T\in \{Q,R,C,N\}$ being sent from participant $x$ to participant $y$ in round $k$. $H_k$ is the set of all messages $M=(T,x,y)$ that have been sent from round $0$ to $k$. $E(y)$ represents the participant whose expertise is needed to complete the task assigned to participant $y$.
\begin{align}
\label{eq: choiceProb} P(M=(T,x,y),k)\propto
\begin{cases}
\langle\theta_{xy}\rangle\cdot\alpha, &\text{if }M\notin H_{k-1}~\wedge~T=Q~\wedge~x\neq y\\
\langle\theta_{xy}\rangle\cdot\beta, &\text{if } M\notin H_{k-1}~\wedge~T=R~\wedge~\{(Q,y,x),(Q,E(y),x)\}\subseteq H_{k-1}\\
\langle\theta_{xy}\rangle\cdot\beta, &\text{if } M\notin H_{k-1}~\wedge~T=C~\wedge~(Q,y,x)\in H_{k-1}~\wedge~E(y)=x\\
\langle\theta_{xy}\rangle\cdot\gamma, &
\text{if }M\notin H_{k-1}~\wedge~T=N~\wedge~(Q,y,x)\in H_{k-1}~\wedge~E(y)\neq x\\
&~~~\wedge~(Q,E(y),x)\notin H_{k-1}\\
0, &\text{otherwise}
\end{cases}
\end{align}

\end{document}